\documentclass[preprint2,12pt]{aastex6}

\usepackage{amsmath}

\def\deg {$^{\circ}$}

\newcommand{\kms}{km\,s$^{-1}$}

\def\ts   {\thinspace}
\def\kms  {\ifmmode{{\rm \ts km\ts s}^{-1}}\else{\ts km\ts s$^{-1}$}\fi}
\def\mo   {\ifmmode{{\rm M}_{\odot}}\else{M$_{\odot}$}\fi}

\def\aco {\ifmmode{^{12}{\rm CO}(J=1\to0)}\else{$^{12}{\rm
CO}(J=1\to0)$}\fi}
\def\bco {\ifmmode{^{12}{\rm CO}(J=2\to1)}\else{$^{12}{\rm
CO}(J=2\to1)$}\fi}
\def\m  {\ifmmode{\mu {\rm m}}\else{$\mu$m}\fi}
\def\cco {\ifmmode{^{13}{\rm CO}(J=1\to0)}\else{$^{13}{\rm
CO}(J=1\to0)$}\fi}
\def\dco {\ifmmode{^{13}{\rm CO}(J=2\to1)}\else{$^{13}{\rm
CO}(J=2\to1)$}\fi}
\def\eco {\ifmmode{{\rm C}^{18}{\rm O}(J=1\to0)}\else{{\rm C}$^{18}{\rm
O}(J=1\to0)$}\fi}
\def\hi  {\ifmmode{{\rm H}{\rm \small I}}\else{H\ts {\scriptsize I}}\fi}
\def\hii  {\ifmmode{{\rm H}{\rm \small II}}\else{H\ts {\scriptsize II}}\fi}
\def\Ha  {\ifmmode{{\rm H}{\alpha}}\else{H\ts {$\alpha$}}\fi}
\def\Hb  {\ifmmode{{\rm H}{\alpha}}\else{H\ts {$\beta$}}\fi}

\def\nh  {\ifmmode{N(\hi)}\else{$N$(\hi)}\fi}
\def\hun  {\ifmmode{I_{100}}\else{$I_{100}$}\fi}
\def\sex  {\ifmmode{I_{60}}\else{$I_{60}$}\fi}
\def\hh   {\ifmmode{{\rm H}_2}\else{H$_2$}\fi}
\def\nhh   {\ifmmode{N({\rm H}_2)}\else{$N$(H$_2$)}\fi}
\def\zwco  {\ifmmode{^{12}{\rm CO}}\else{$^{12}{\rm CO}$}\fi}
\def\nzwco  {\ifmmode{N(^{12}{\rm CO})}\else{$N(^{12}{\rm CO})$}\fi}
\def\wzwco  {\ifmmode{W(^{12}{\rm CO})}\else{$W(^{12}{\rm CO})$}\fi}
\def\drco  {\ifmmode{^{13}{\rm CO}}\else{$^{13}{\rm CO}$}\fi}
\def\ndrco  {\ifmmode{N(^{13}{\rm CO})}\else{$N(^{13}{\rm CO})$}\fi}
\def\wdrco  {\ifmmode{W(^{13}{\rm CO})}\else{$W(^{13}{\rm CO})$}\fi}
\def\tex  {\ifmmode{T_{ex}({\rm CO})}\else{$T_{ex}({\rm CO})$}\fi}
\def\xco   {\ifmmode{X_{\rm CO}}\else{$X_{\rm CO}$}\fi}
\def\msol   {\ifmmode{{\rm M}_{\odot}}\else{M$_{\odot}$}\fi}
\def\amm    {NH$_{3}$}

\def\ha     {H$\alpha$}
\def\Pa     {Pa$\alpha$}

\def\methanol {CH$_3$OH}
\def\water {H$_2$O} 
\def\sfr {M$_{\odot}$\,yr$^{-1}$}
\def\Lsun{L$_{\odot}$}
\def\Msun{M$_{\odot}$}
\def\cn {CN(1-0;1/2-1/2)}

\submitted{Accepted for publication in the Astrophysical Journal}

\begin{document}

\title{Survey of Water and Ammonia in Nearby galaxies (SWAN): Resolved Ammonia Thermometry, Water and Methanol Masers in the Nuclear Starburst of NGC 253}

\author{Mark Gorski\altaffilmark{1,2}}
\email{mgorski@unm.edu}

\author{J\"urgen Ott\altaffilmark{1}}
\email{jott@nrao.edu}

\author{Richard Rand\altaffilmark{2}}
\email{rjr@unm.edu}

\author{David S. Meier\altaffilmark{3,1}}
\email{david.meier@nmt.edu}

\author{Emmanuel Momjian\altaffilmark{1}}
\email{emomjian@nrao.edu}

\author{Eva Schinnerer\altaffilmark{4}}
\email{schinner@mpia.de}

\affil{1 National Radio Astronomy Observatory, P.O. Box O, 1003 Lopezville Road, Socorro, NM 87801, USA}
\affil{2 Department of Physics and Astronomy, University of New Mexico, 1919 Lomas Blvd NE, Albuquerque, NM 87131, USA}
\affil{3 Department of Physics, New Mexico Institute of Mining and Technology, 801 Leroy Place, Socorro, NM 87801, USA}
\affil{4 Max-Planck Institut f\"ur Astronomie, K\"onigstuhl 17, D-69117 Heidelberg, Germany}


\begin{abstract}

We present Karl G Jansky Very Large Array molecular line observations of the nearby starburst galaxy NGC\,253, from SWAN: "Survey of Water and Ammonia in Nearby galaxies". SWAN is a molecular line survey at centimeter wavelengths designed to reveal the physical conditions of star forming gas over a range of star forming galaxies. NGC\,253 has been observed in four 1GHz bands from 21 to 36\,GHz at 6\arcsec $\sim100$\,pc) spatial and 3.5\,\kms spectral resolution. In total we detect 19 transitions from seven molecular and atomic species. We have targeted the metastable inversion transitions of ammonia (NH$_{3}$) from (1,1) to (5,5) and the (9,9) line, the 22.2\,GHz water (H$_2$O) ($6_{16}-5_{23}$) maser, and the 36.1\,GHz methanol (CH$_3$OH) ($4_{-1}-3_{0}$) maser. Utilizing NH$_{3}$ as a thermometer, we present evidence for uniform heating over the central kpc of NGC\,253. The molecular gas is best described by a two kinetic temperature model with a warm 130K and a cooler 57K component. A comparison of these observations with previous ALMA results suggests that the molecular gas is not heated in photon dominated regions or shocks. It is possible that the gas is heated by turbulence or cosmic rays. In the galaxy center we find evidence for NH$_{3}$(3,3) masers. Furthermore we present velocities and luminosities of three water maser features related to the nuclear starburst. We partially resolve CH$_3$OH masers seen at the edges of the bright molecular emission, which coincides with expanding molecular superbubbles. This suggests that the masers are pumped by weak shocks in the bubble surfaces.

\end{abstract}


\section{Introduction}

Physical descriptions of how the Interstellar Medium (ISM) condenses to form stars, and the resulting feedback from star formation, limit our understanding of galaxy evolution and the star formation history of the universe.  Models of galaxy evolution without feedback drastically over predict star formation rates and efficiencies. Feedback is necessary to impede star formation otherwise a galaxy reach its peak star formation rate in less than a dynamical time and quickly convert its baryons to stars shortly after (e.g. \citealt{Kauffmann1999}; \citealt{Krumholz2011}; \citealt{Hopkins2011}). Star formation is largely correlated with the amount of dense, $\geq10^4$ cm$^{-3}$, molecular gas within a galaxy (e.g. \citealt{Gao2004}), and in simulations the state of such gas is limited by the resolution of each individual simulation, i.e. subgrid physics 
(e.g. \citealt{Okamoto2005}; \citealt{Haas2013}; \citealt{Crain2015}).  Consequently, understanding the relationship between star formation, the resulting feedback, and the state of the dense molecular gas is critical to understanding the star formation history of a galaxy. 

Feedback in the star forming ISM perturbs {molecular gas} such that it can no longer collapse and form new stars. Heating from supernovae, stellar winds, and photoionization are examples of the most dominant forms of stellar feedback (e.g. \citealt{Kauffmann1999}, \citealt{Hartmann2001}, \& \citealt{Vaz2010}). One dimensional simulations by \citet{Murray2010} suggest that different mechanisms dominate at different times during the lifetime of  star forming Giant Molecular Clouds (GMCs). Furthermore, simulations by \citet{Hopkins2012}  and \citet{Hopkins2014} show that these different feedback effects compound in a non-linear way and that no single feedback process dominates {the star forming ISM}. Additionally, the environment in which {star forming material exists} may play a critical role in its properties. Therefore, observational constraints are necessary to refine the subgrid physics within the theoretical models on the appropriate scales.

Nearby galaxies provide access to scales of tens of pc where observations can reveal how feedback operates.  There are many studies that look at different aspects of feedback. For example in NGC 253 \citet{Strick2002} and \citet{Westm2011} discuss the X-ray and ionized gas properties of a starburst driven outflow, respectively. Studies of \amm, such as those by \citet{Ott2005}, \citet{Lebron2011} and \citet{Mangum2013}, reveal heating and cooling of the molecular ISM. In addition, other molecular tracers reveal shocks, Photon Dominated Regions (PDRs), masses, lengths and time scales associated with star formation. (e.g. \citealt{Meier2015}, \citealt{leroy2015}). 

The ``Survey of Water and Ammonia in Nearby galaxies" (SWAN) is a survey of molecular line tracers at centimeter wavelengths designed to reveal the physical conditions in star forming gas.
The  sample consists of four star forming galaxies: NGC 253, IC 342, NGC 6946, and NGC 2146, and was chosen to span a {range of galaxy types} from Milky Way-like to starbursts and an order of magnitude of star formation rates from ($\lesssim1$ \sfr) to starbursts ($\sim$10 \sfr).
Here we discuss the first results from SWAN focusing on VLA  K- and Ka-band observations of NGC 253. 

\begin{figure}
\centering
\includegraphics[width=0.5\textwidth]{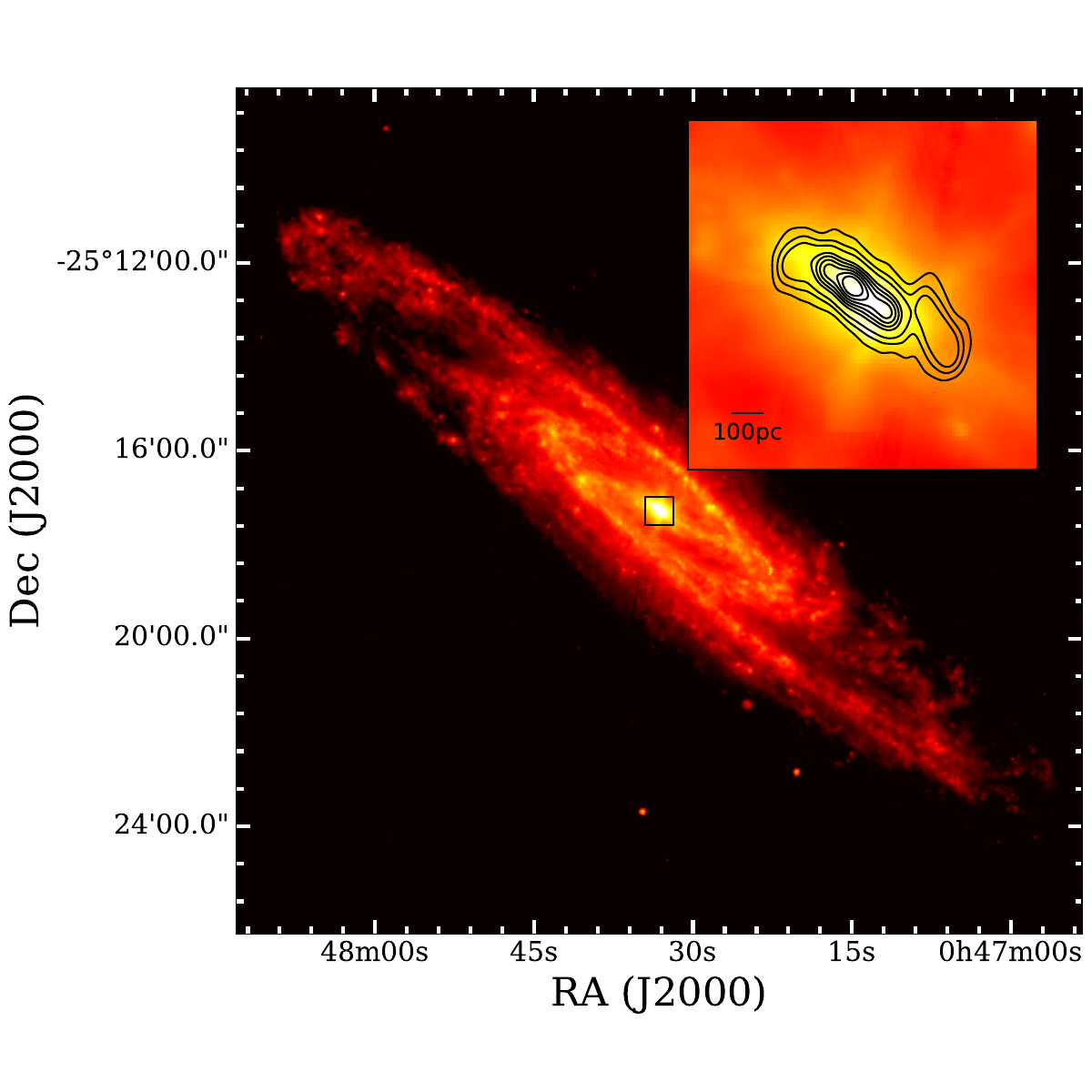}
\caption{Spitzer IRAC 8.0$\mu$m image of NGC 253 \citep{Dale2009}. The inset shows the central kpc with \amm(3,3) 3, to 30$\sigma$ contours, with steps of 3$\sigma$, showing the dense molecular gas associated with the nuclear starburst. The 3$\sigma$ contour equates to 4.7 mJy beam$^{-1}$ \kms.}
\label{fig:irac}
\end{figure}

NGC 253 has been studied at many wavelengths: X-ray (e.g. \citealt{Strick2002}), optical (e.g. \citealt{Westm2011}),  infrared (e.g. \citealt{Dale2009}), millimeter (e.g. \citealt{BA2013} \& \citealt{Meier2015}), and radio (e.g. \citealt{UA1997}). Figure \ref{fig:irac} shows a Spitzer 8$\mu$m image of NGC 253 \citep{Dale2009}. The inset shows the central kpc with 3, 6 and 9$\sigma$ contours of \amm(3,3) emission from our data discussed in Section 3.2.1.  We adopt a distance of 3.5 Mpc measured from the tip of the red giant branch \citep{Radburn-Smith} and a systemic velocity of 234 \kms\ in the LSRK frame from \citet{W1999}. All velocities in this paper will be in the LSRK frame unless otherwise stated.  The disk is inclined at i$\sim$78\deg\ \citep{Pence1980}. NGC 253 has a total star formation rate (SFR) of $\sim$5.9 \sfr\, of which approximately half is concentrated into the central kpc \citep{McCormick2013}.  The starburst is driving a massive molecular outflow, with a mass-loss rate estimated at 9\,\sfr, that is thought to be starving the current star forming event \citep{BA2013}. The outflow is also seen in X-rays \citep{Strick2002} and \ha\ \citep{Watson1996}. \citet{Saka2006} found evidence for two expanding molecular superbubbles within the central kpc with kinetic energies of order $\sim10^{46}$ J. \citet{Ott2005} and \citet{BA2013} found several smaller molecular superbubbles in the same region. \citet{BA2013} suggest that superbubbles and supernovae from the starburst drive the wind, whereas \citet{Westm2011} favor a cosmic ray driven wind on the larger scales with a small contribution from the starburst in the center, with the molecular gas in the center responsible for collimating the outflow. 

Centimeter and millimeter wavelength spectra of galaxies provide access to diagnostically important molecular tracers. The molecular gas is often well traced by CO, however more complex molecules can provide better tracers of gas properties such as temperature and density, and can be used to trace specific conditions such as PDRs and shocks (e.g. \citealt{Fuente1993}, \citealt{GB2000}, \citealt{MT2005}, and \citealt{Meier2015}). This study will focus on the metastable transitions (J=K) (1,1) to (5,5) and (9,9) of \amm,  the 22 GHz \water($6_{16}-5_{23}$) maser, and the 36 GHz \methanol($4_{14}-3_{03}$) maser.  We will use a combined analysis of these lines to expose the processes that dominate the central kpc of NGC 253.
 
The \amm\,molecule generally works well as a temperature tracer of the molecular gas. The tetrahedral structure of \amm\ makes it a symmetric top, meaning the energy states are described by the rotation angular momentum quantum number J and the projection along the symmetry axis K. The J=K states, called metastable states, are long lived compared to J$>$K states, and population exchanges between K ladders are forbidden except by collisions. Therefore when \amm\ is collisionally excited ( critical density n$_{H_2}\gtrsim10^3$ cm$^{-3}$) the K ladders are expected to be populated in accordance with the kinetic temperature of the gas.  As a result, measurements of the relative intensities of metastable states act as probes of the rotation temperature of the gas (e.g. \citealt{H&T}, \citealt{walmsley83}, \citealt{Lebron2011}, \citealt{Ott2005}, \citealt{OH2011}, and \citealt{Mangum2013}).

 The \amm(3,3) state can be a maser transition \citep{walmsley83}, but it is less well studied than other masers. In the Galaxy there is a weak association of \amm(3,3) masers with dense gas in star forming regions (e.g. \citealt{Wilson1990}, \citealt{Mills2013}, and \citealt{Goddi2015}). Here we will use metastable transitions of \amm\ to understand the heating and cooling balance of the dense molecular ISM in NGC 253. 

The \water\ and \methanol\ masers provide a unique opportunity to probe star forming environments. The \water\ line requires gas densities $>10^7\,$cm$^{-3}$ and kinetic temperatures $>300$ K to mase (e.g. \citealt{Tarchi2012}). In the Galaxy these masers are typically found in shocked regions around Young Stellar Objects (YSOs) and Asymptotic Giant Branch (AGB) stars (e.g. \citealt{Palagi1993}) and may be used to identify regions of hot, dense, and/or shocked gas. This is in addition to precisely tracing kinematics of stellar winds (e.g. \citealt{Goddi2006}), and accretion disks (e.g. \citealt{Peck2003}; \citealt{Lo2005}; \citealt{Reid2009}). 

Class I (collisionally pumped) and II (radiatively pumped)  \methanol\ masers are found in high mass star forming regions (e.g. \citealt{Ell2012}) and supernova remnants (e.g. \citealt{McEwen2014})in the Galaxy. The Class I masers trace shocks and gas densities $>10^4\,$cm$^{-3}$ \citep{pratap2008}. The 36.2 GHz \methanol\ line studied here is a Class I type maser. We will mostly use these masers as signposts of shocked material. 

In \S 2 we describe the observational setup. In \S 3 we report our measurements of the \amm, \water, and \methanol\ lines in addition to a brief description of the continuum and the H56$\alpha$ Radio Recombination Line (RRL). In \S 4 we discuss the derivation of temperatures across the molecular bar, the relevance of the \water\ masers to the outflow, the significance of the \methanol\ masers, and a comparison with previous ALMA millimeter molecular lines. Lastly, we summarize our findings in \S 5.


\section{Observations and Data Reduction}

We observed NGC 253 with the 18-26.5 GHz and 26.5-40 GHz (K- and Ka-bands) receivers of the Karl G. Jansky Very Large Array (VLA)\footnote{The National Radio Astronomy Observatory is a facility of the National Science Foundation operated under cooperative agreement by Associated Universities, Inc.} (project code: 13A-375). The K-band observations were carried out on 2013 May 11. The Ka-band observations were split into two sessions:  2013 May 23 and 2013 May 26.  The VLA was in the DnC hybrid configuration for all these observations. This configuration delivers a rounder beam for low declination sources because the north arm is in the more extended C configuration, while the east and west arms are in D configuration. The received signal is sampled at each antenna using the 8-bit samplers. These provide two 1 GHz baseband pairs with both right and left hand circular polarizations. The correlator was set up to divide each baseband into 8 sub-bands each with 512 channels resulting in a channel width of 250kHz. This yields a velocity resolution ranging from 3.0 to 3.3\kms\ for the K-band,  and 2.0 to 2.7 \kms\ for the Ka-band observations.  The baseband pairs were centered at 21.8 GHz and 24.1 GHz in K-band, 27.1 GHz and 36.4 GHz in Ka-band. They will be referred to as the 22 GHz, 24 GHz, 27 GHz, and 36 GHz basebands, respectively. These were chosen to include the metastable \amm\ transitions from (1,1) with rest frequency 26.6946 GHz, to (5,5) at 24.5330 GHz, as well as (9,9) at 27.4779 GHz, the \water($6_{16}-5_{23}$)  maser line at 22.2351 GHz, and the \methanol($4_{14}-3_{03}$) transition at 36.1693 GHz. All of the detected lines and their rest frequencies are listed in Table \ref{tab:Lines}.  The on source time for the K-band and Ka-band observations was 4.4 hours and 3.6 hours, respectively.  We used 3C48 as the flux density calibrator 
\cite{PB2013}, J2253+1608 as the bandpass calibrator, and J0025-2602 as the complex gain calibrator in all observations. We alternated between 10 minute intervals on NGC 253 and 1.5 minute intervals on the complex gain calibrator. 

The data were reduced in the Common Astronomy Software Applications (CASA) package version 4.2.2 \citep{mcmullin07}.  At the adopted distance of 3.5 Mpc the linear scale is $\sim$17 pc per arcsecond. The half power primary  beam widths of the VLA for the K- and Ka-bands are 2.1\arcmin\ ($\sim$2 kpc) and 1.5\arcmin\ ($\sim$1.5 kpc) respectively.  All data cubes were gridded with 0.25\arcsec\ pixels, mapped using natural weighting, CLEANed to $\sim3\sigma$ rms noise, and are regridded to a common velocity resolution of 3.5\kms. Continuum subtraction was performed in the UV domain. For the 24 GHz baseband, several baselines were flagged for being noisy yielding a slightly larger synthesized beam than the 22GHz baseband. The resulting image cubes are then smoothed to a common synthesized beam of 6\arcsec$\times$4\arcsec\ (Position Angle: 3.00\deg). The common resolution cubes are used for consistency for all the observed lines. The resulting rms noise values in the K-band and Ka-band image cubes are 0.5 mJy beam$^{-1}$  and 1 mJy beam$^{-1}$ in a 3.5 \kms\ channel, respectively. The maser lines of \water, \methanol, and \amm(3,3) have also been imaged with 0.25\arcsec\ pixels and Briggs (robust=0) weighting, yielding a synthesized beam of 4\arcsec$\times$3\arcsec\ for \water\ and \amm(3,3) and 2\arcsec$\times$1\arcsec\ for \methanol, to better constrain the locations of masing material.  The rms noise in the K-band and Ka-band Briggs weighted image cubes are 1.5 mJy beam$^{-1}$ and 3.1 mJy beam$^{-1}$ in a 3.5 \kms\ channel, respectively.  A super-resolved cube was constructed for the \water\ maser data. This was done by deconvolving a dirty image cube then convolving the CLEAN components with a 1\arcsec\ round beam. This super-resolved image cube is used to emphasize structures seen {in the CLEAN components that are
difficult to discern in the regularly resolved image cubes.} No quantitative measurements are made with the super-resolved cube.  


\begin{deluxetable}{lc}
\tablecaption{Detected Molecular and Atomic Transitions\label{tab:Lines}}
\tabletypesize{\scriptsize}
\tablewidth{0pt}
\tablehead{
\colhead{Transition} & \colhead{Rest Frequency} \\
	& (GHz)  
}
\startdata
c-C$_3$H$_2$ \,(2$_{20}$-2$_{11}$)	& 21.5874 	\\
\textbf{\water($6_{16}-5_{23}$) }			& \textbf{22.2351} 	\\
H66$\alpha$		& 22.3642 	\\
H64$\alpha$ 		& 24.5099 	\\
\textbf{\amm\,(1,1)} 		& \textbf{23.6945} 	\\
\textbf{\amm\,(2,2)}		&\textbf{23.7226}	\\
\textbf{\amm\,(3,3)}		& \textbf{23.8701}		\\
\textbf{\amm\,(4,4)} 		& \textbf{24.1394} 	\\
\textbf{\amm\,(5,5) }		& \textbf{24.5330}	\\
H62$\alpha$			& 26.9392	\\
c-C$_3$H$_2$ \,(3$_{30}$-3$_{21}$) 	& 27.0843 \\
HC$_3$N\,(3-2)	& 27.2944 \\
\textbf{\amm\,(9,9)}	&	\textbf{27.4779}	\\
\textbf{\methanol\,($4_{-1}-3_{3}$)}	&	\textbf{36.1693}	\\
HC$_3$N\,(4-3)	& 36.3924	 \\
H56$\alpha$			& 36.4663 \\
CH$_3$CN\,(2-1)	& 36.7956 \\
\enddata
\tablecomments{Molecules selected for this paper's analysis are shown in bold face text}
\end{deluxetable}



\section{Results}

\begin{figure}
\includegraphics[width=0.49\textwidth]{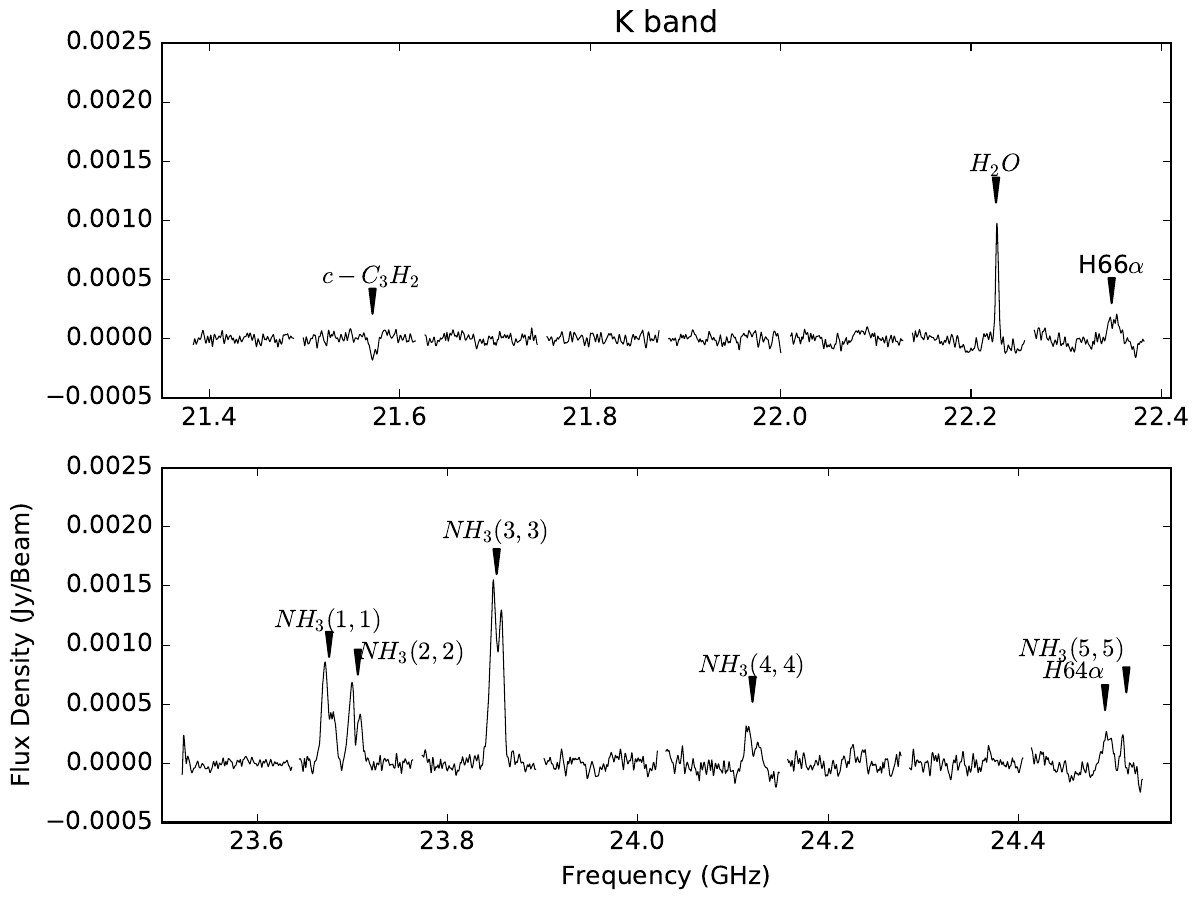}
\includegraphics[width=0.49\textwidth]{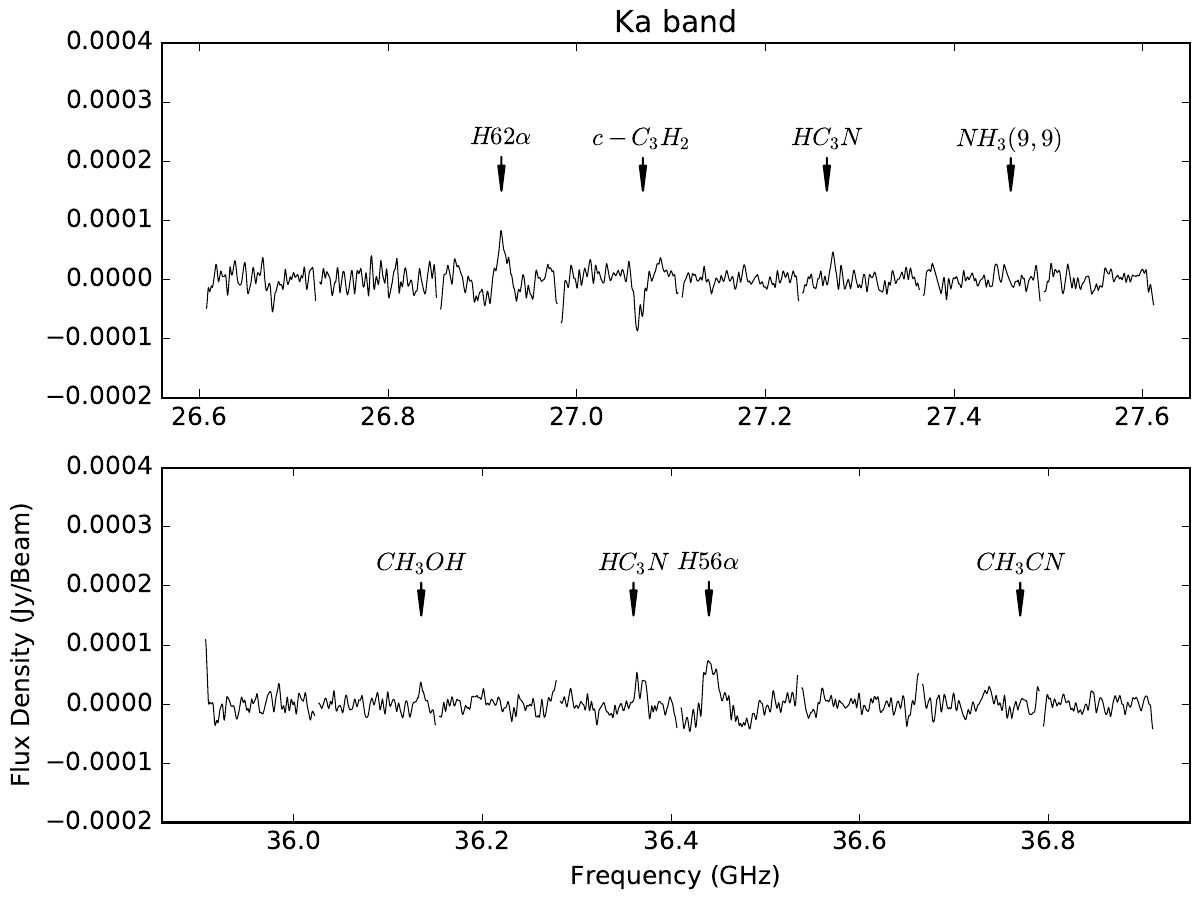}
\caption{The observed spectrum from K-band and Ka-band. The spectrum was extracted from a 40\arcsec$\times$25\arcsec\ box centered at RA: 00$^h$47$^m$33.12$^s$ DEC: -25$^\circ$17'19.33" and Hanning smoothed with a window of 21 channels . Each box shows the selected 1GHz basebands with frequency on the X-axis and flux density on the Y-axis. Detected molecular and atomic species are identified with black arrows. }
\label{fig:spectrumfull}
\end{figure}

The full continuum subtracted spectra of our K- and Ka-band observations are presented in Figure~\ref{fig:spectrumfull}. We have identified 17 transitions from seven different molecular or atomic species shown in Table \ref{tab:Lines}. Lines were identified by searching the common resolution data cubes with a pixel sized beam. We set conservative conditions for a detection at a peak flux of $>$1.5 mJy beam$^{-1}$ channel$^{-1}$ for K-band and $>$3.0 mJy beam$^{-1}$ channel$^{-1}$ for Ka-band, and a FWHM of $\gtrsim$30 \kms\ for thermal lines. For known maser transitions single channels above 6$\sigma$ are considered detections. 


\subsection{Continuum Emission and Radio Recombination Lines}

\begin{figure*}
\includegraphics[width=0.9\textwidth]{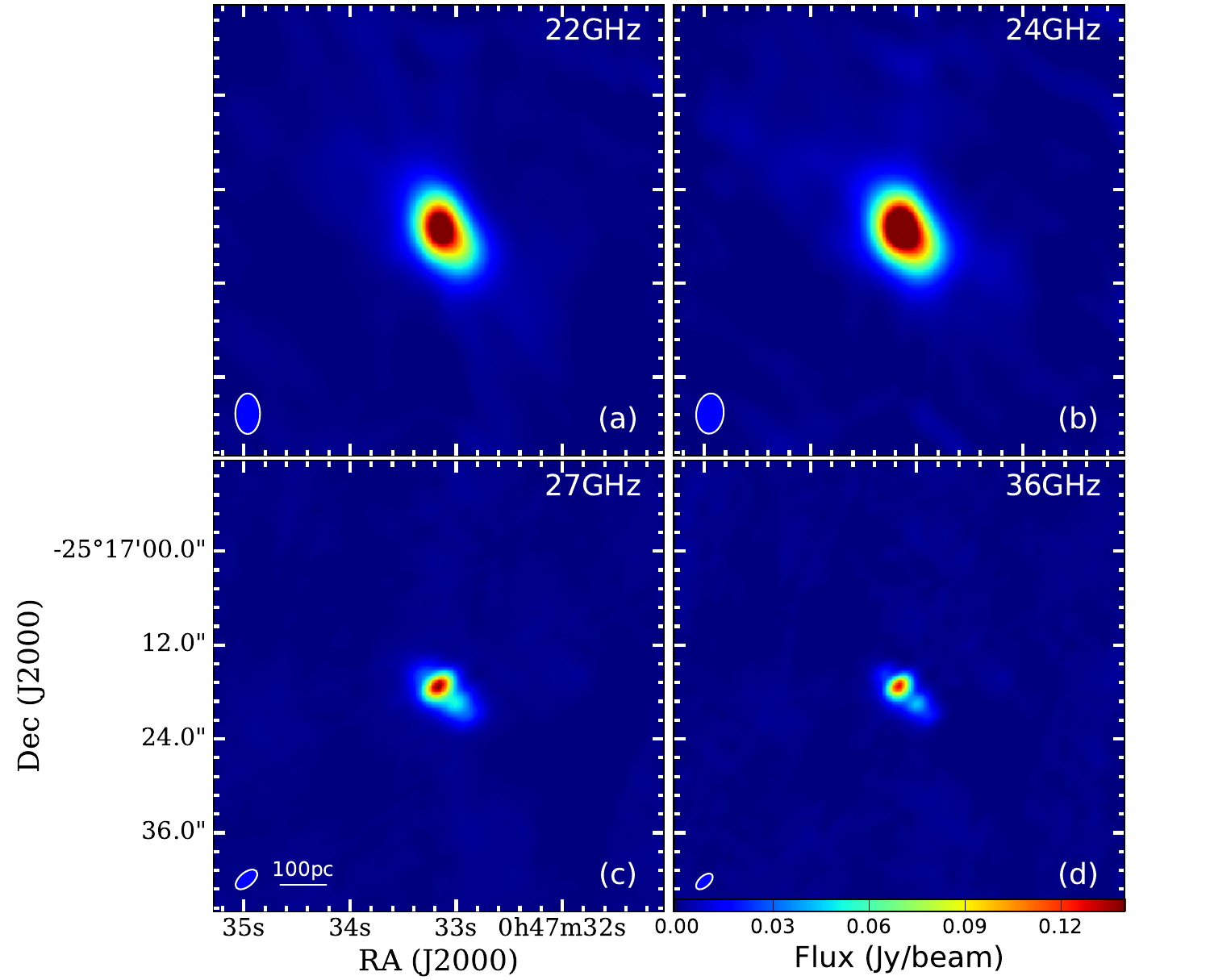}
\caption{Images of the continuum emission from each of the 1 GHz basebands. The images are Briggs weighted and have not been smoothed to the common $6\arcsec \times 4\arcsec$ resolution. }
\label{fig:cont}
\end{figure*}

\begin{figure*}
\includegraphics[width=0.9\textwidth]{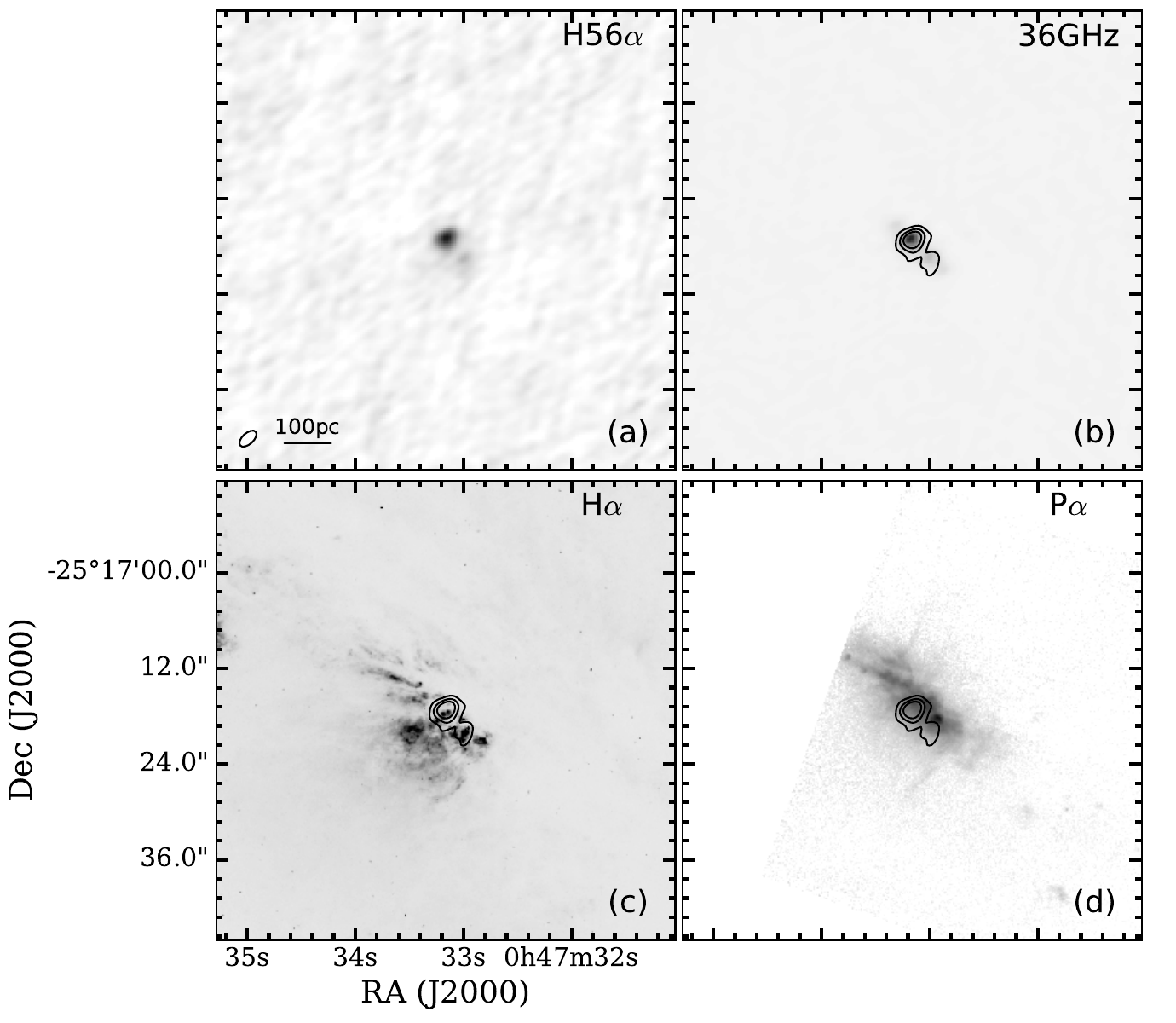}
\caption{Comparison of star formation and outflow tracers. The contours show the 3, 6, and 9 $\sigma$ contours of the naturally weighted H56$\alpha$ RRL where 1$\sigma$ is 0.08 Jy beam$^{-1}$ \kms. (a) H56$\alpha$ RRL shows the least dust obscured star formation. (b) the 36 GHz continuum image shows a close correlation with the two knots in the RRL. (c)  HST WFPC2 H$\alpha$ image \citep{Watson1996}.  (d) HST P$\alpha$ \citep{AH2003} shows the outflow less well than the H$\alpha$ but is less obscured by dust.}
\label{fig:SFmap}
\end{figure*}

The radio continuum emission from NGC 253 is resolved in all four basebands (Figure \ref{fig:cont}). Images were made selecting only line free channels. The flux density measured in the 24 GHz baseband is 550$\pm$30 mJy. This agrees with the value of 520$\pm$52 mJy from \cite{Ott2005}. The 36 GHz flux density measures 350$\pm$40 mJy in close agreement with the value of 330 mJy at 32 GHz measured by \citet{Kepley2011}. The other continuum flux density measurements are 470$\pm$20 mJy and 370$\pm$40 mJy for the 22 GHz and 27 GHz basebands, respectively. The spectral index derived from the K-band basebands is  $-1.8\pm$0.5 and for Ka-band $0.2\pm0.4$. The measurements suggest the continuum goes through a minimum between the 24 GHz and 27 GHz basebands.  The H56$\alpha$ transition is the strongest radio recombination line (RRL) we detect. Figure \ref{fig:SFmap} shows the relationship between the Hubble (HST) \ha\ \citep{Watson1996}, H56$\alpha$, \Pa\ \citep{AH2003}, and the 36 GHz continuum emission. We treat the H$\alpha$ as a tracer of the outflow with heavy dust obscuration, \Pa\ as a partially obscured star formation tracer and RRL as an unobscured star formation tracer. 

\subsection{Molecular Emission Lines}

\begin{figure*}
\includegraphics[width=0.95\textwidth]{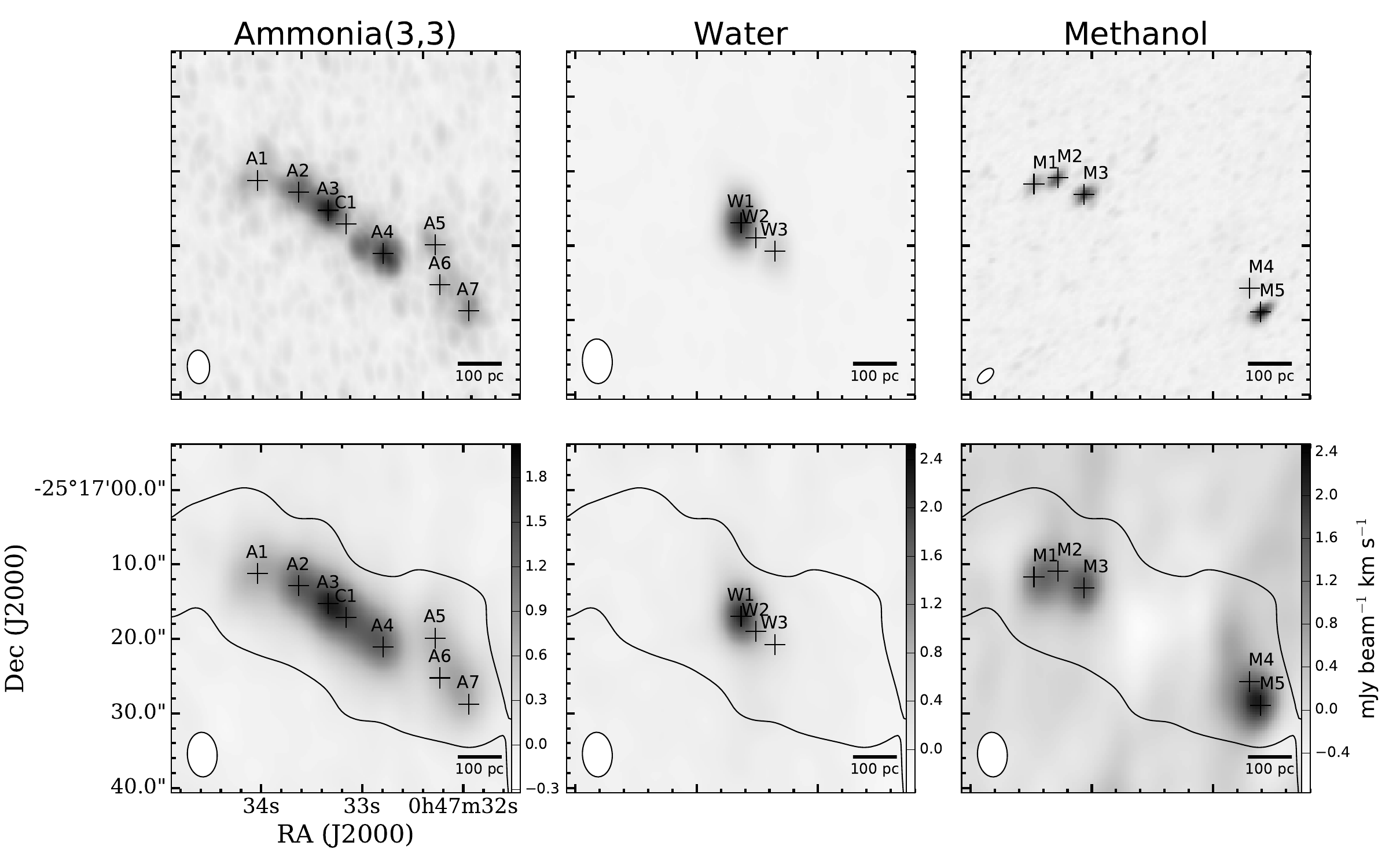}
\caption{ Images of the \amm(3,3) (left column), \water\ (center column), and \methanol\ (right column) lines. The top row shows the peak flux images of the unsmoothed naturally weighted image cubes spanning 300\kms about systemic velocity. The bottom row shows the intensity images smoothed to the common resolution of $6\arcsec\times4\arcsec$. Both the peak flux and intensity images are plotted with the same greyscale. The crosses mark locations where spectra were extracted for analysis, and the continuum peak is marked C1. The contour is  60 Jy beam$^{-1}$\kms  of \aco\ from \citet{BA2013} smoothed to match the resolution of the VLA data.} 
\label{fig:integrated}
\end{figure*}

Our analysis from here on will focus only on \amm, \water, and \methanol. These transitions are shown in bold text in Table \ref{tab:Lines}.  Intensity and peak flux maps are shown in Figure \ref{fig:integrated}. Spectra have been extracted from the naturally weighted, smoothed cubes from the pixels at the locations marked by crosses (Figure \ref{fig:integrated}) and  are shown in Figures \ref{fig:linemos}, \ref{fig:waterspec},  and \ref{fig:methanolspec}. These locations were selected from the spatial peaks in the peak flux maps shown, or for spectrally unique characteristics. For example,  W2 is not a spatial peak in the peak flux map but was selected for a spectral component discussed in Section 3.2.2. The continuum peak is labeled C1, and identifies the starburst center. The \amm\ locations are labeled 
A1-A7. We use the \amm(3,3) line to select a representative sample of locations across NGC 253 because it is the strongest observed \amm\ transition. Locations for \water\,and \methanol\ are labeled with W and M, respectively.


\subsubsection{\amm\,Inversion Lines}

\begin{figure*}
\centering\includegraphics[width=0.95\textwidth]{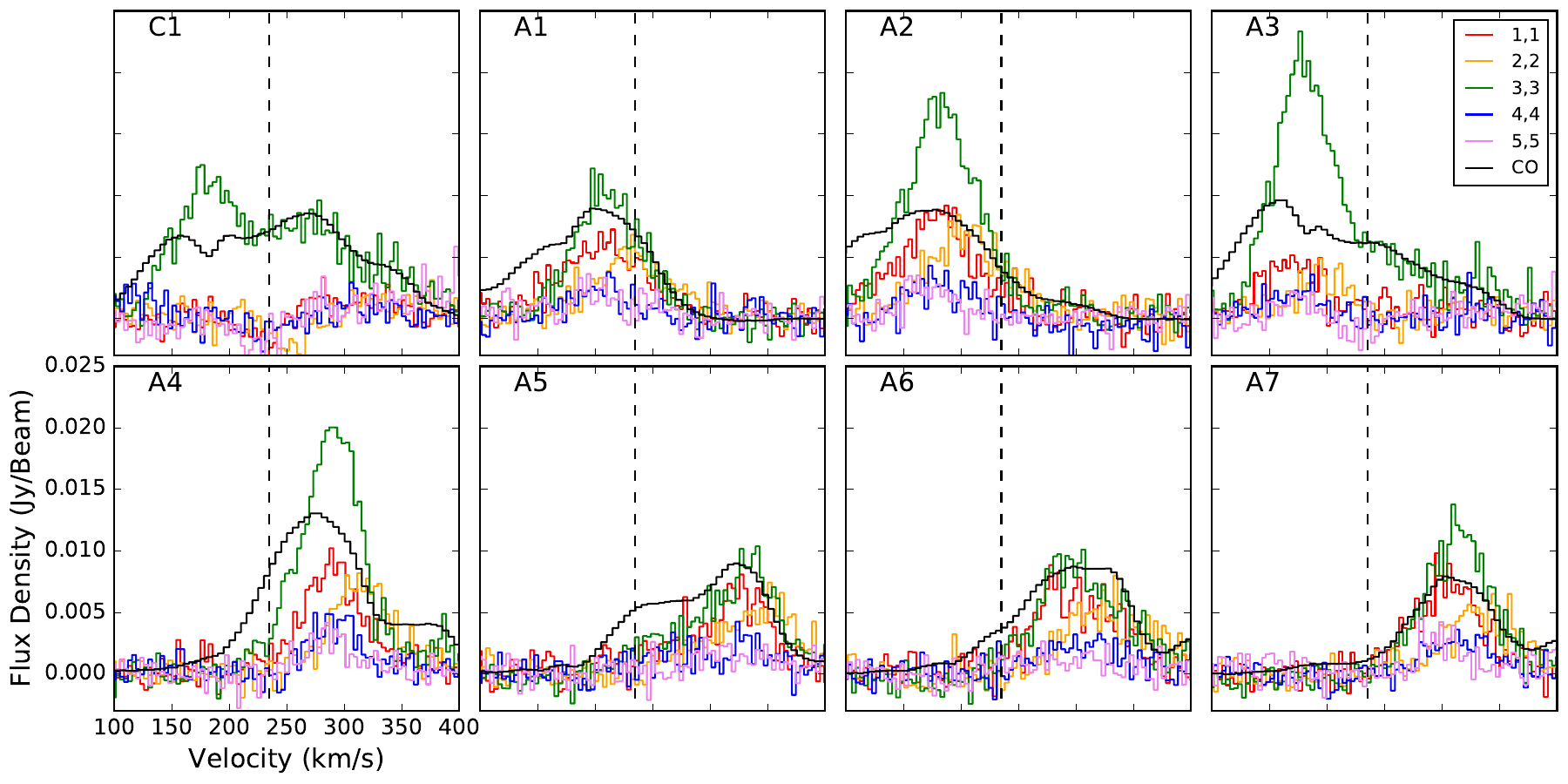}
\caption{Spectra of the \amm(1,1) to (5,5) transitions extracted from locations A1-A7 and C1. \amm(3,3), plotted in green, shows emission at all locations, whereas all the other inversion lines are absorbed toward C1. We have plotted the \aco\ profile (flux density scaling factor of 0.002 and 2.5\kms\ resolution) from \citet{BA2013} in black for comparison. The vertical dashed line denotes the systemic velocity of NGC 253 of 234 \kms.}
\label{fig:linemos}
\end{figure*}

\begin{figure*}
\centering\includegraphics[width=0.8\textwidth]{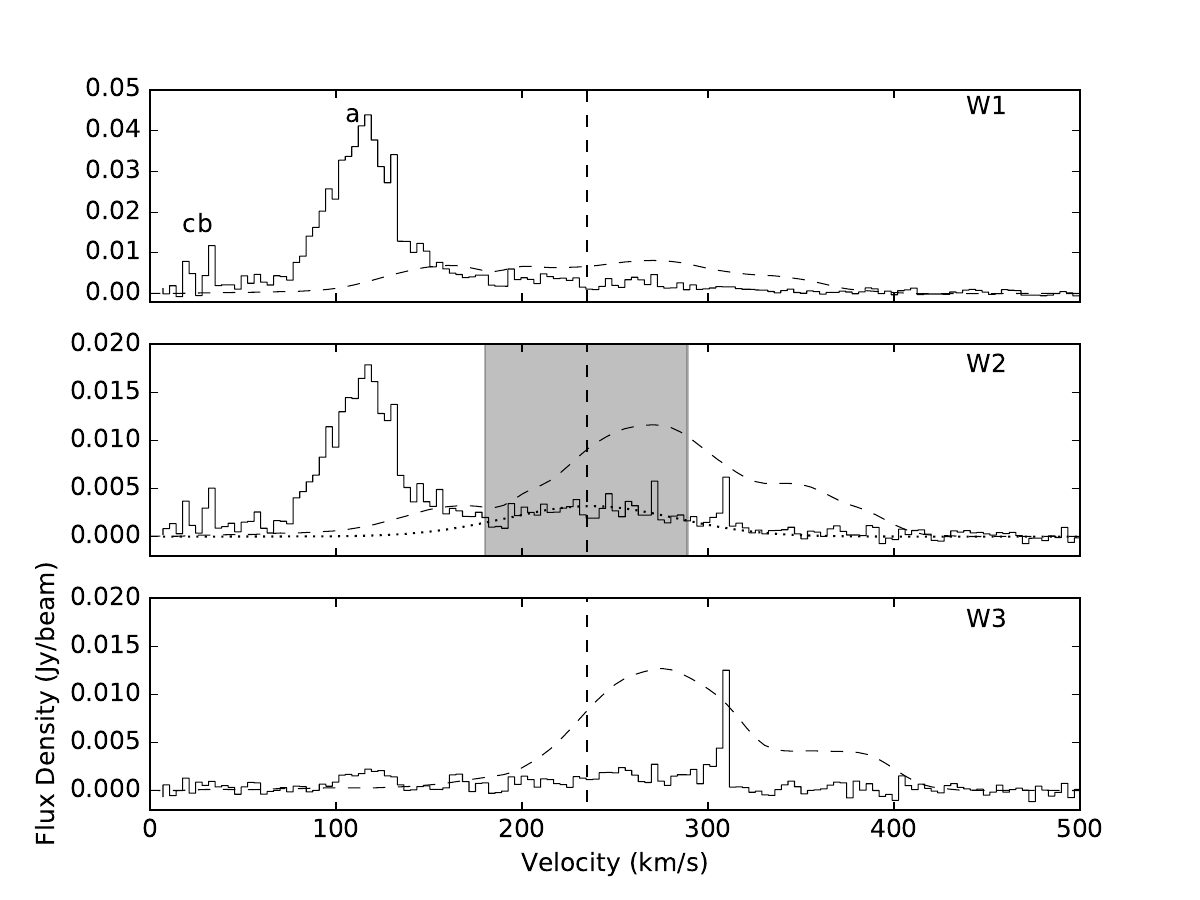}
\caption{Water maser spectra from W1, W2, and W3. W1 shows multiple velocity components. These are labeled a to c. W2 is a 
spectrally broad component centered at 233 \kms. It is highlighted by the grey box and the best fit spectrum is shown by the dotted line. This component appears to consist of many individual masers. The dashed line shows the CO spectrum, with 2.5\kms\ resolution, at each location scaled by 0.005 in flux density, from \cite{BA2013}.  }
\label{fig:waterspec}
\end{figure*}

\begin{figure*}
\centering\includegraphics[width=1.0\textwidth]{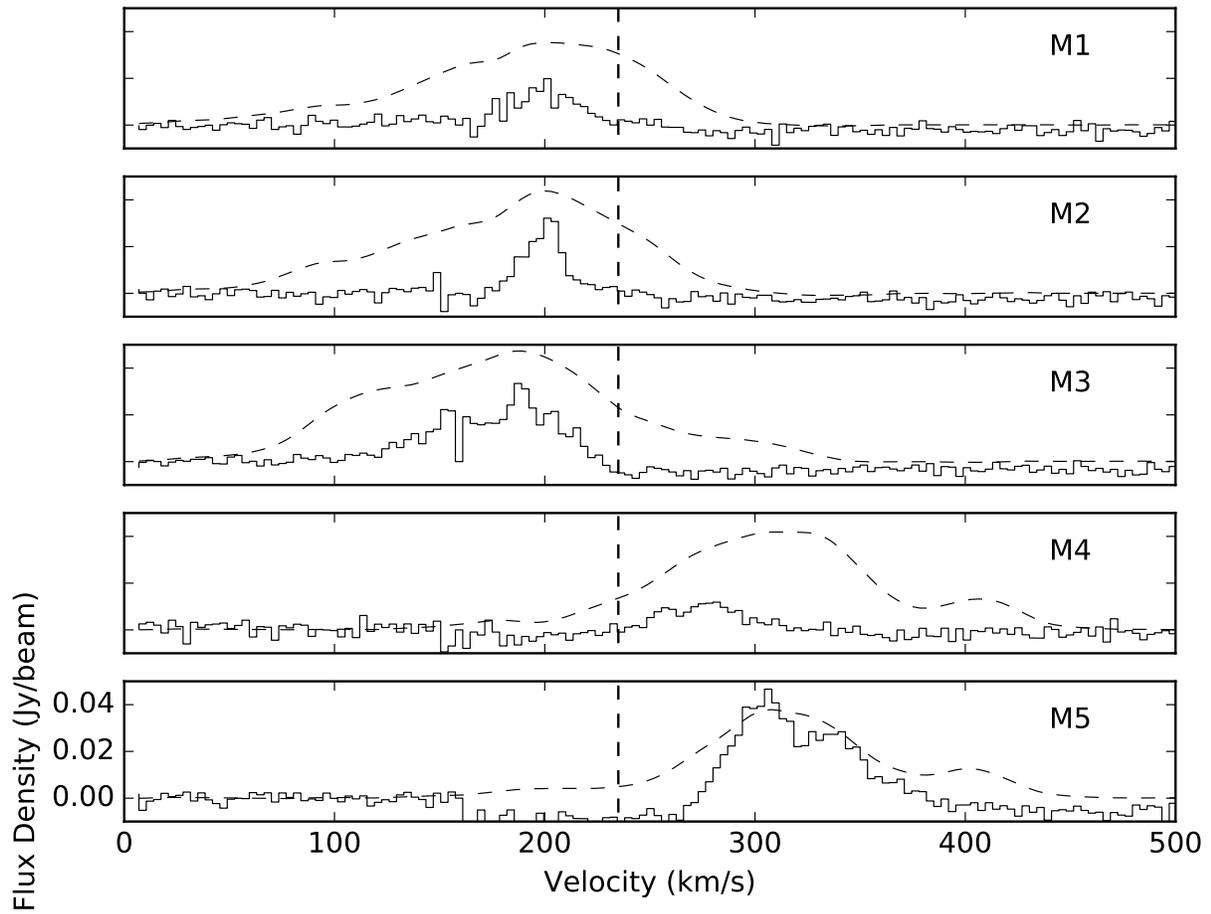}
\caption{Methanol Spectra from M1 to M5. CO spectra are plotted with a dashed line with a flux density scaling factor of 0.01. The systemic velocity of NGC 253 is denoted with a vertical dashed line at 234 \kms.}
\label{fig:methanolspec}
\end{figure*}

The observed \amm\ emission spans an elongated structure $\sim$1 kpc in length about the continuum peak. The northeast (NE) side of the continuum peak contains three \amm(3,3) spatial peaks (A1, A2 \& A3), {which are blueshifted} from the systemic velocity with observed velocities ranging from $\sim160$\kms\ to 200\kms.The southwest (SW) side of the continuum peak contains four \amm(3,3) peaks (A4, A5, A6, \& A7), {which are redshifted} from systemic with velocities ranging from $\sim280$\kms\ to 320\kms. (Figure \ref{fig:integrated}).  
 The peaks A1, A2, \& A3 can be cross-identified with in regions F, E, and D from \citet{Ott2005} and A4, A5, A6 and A7 are analogous to C, B, and A . It should be noted that this is not a one to one mapping as we have a smaller beam than the superresolved cube used in \citet{Ott2005}.
 The \amm\ spectra from each pixel are shown in Figure \ref{fig:linemos}. We detect inversion transitions \amm(1,1) to (5,5) at all locations. In addition the \amm(9,9) line is weakly detected at site A3 (spectrum not shown). At the continuum peak all the \amm\ transitions are seen in absorption with the exception of \amm(3,3). Single Gaussians were fitted to the spectrum extracted from each location, from which we extract the integrated flux, peak flux, FWHM, and the line center, from the individual pixels marked in Figure \ref{fig:integrated}. The properties of the metastable transitions, \amm(1,1) to (5,5),  for A1-A7 are listed in Table \ref{tab:ammprop} and C1 in Table \ref{tab:C1prop}. The weakly detected \amm(9,9) fit results are listed in Table \ref{99prop}. We do not see the \amm\ metastable inversion hyperfine transitions (J=K, $\Delta$F = 1) as the lines are likely weak and broad enough to be smeared out. 

\begin{figure}
\centering
\includegraphics[width=0.49\textwidth]{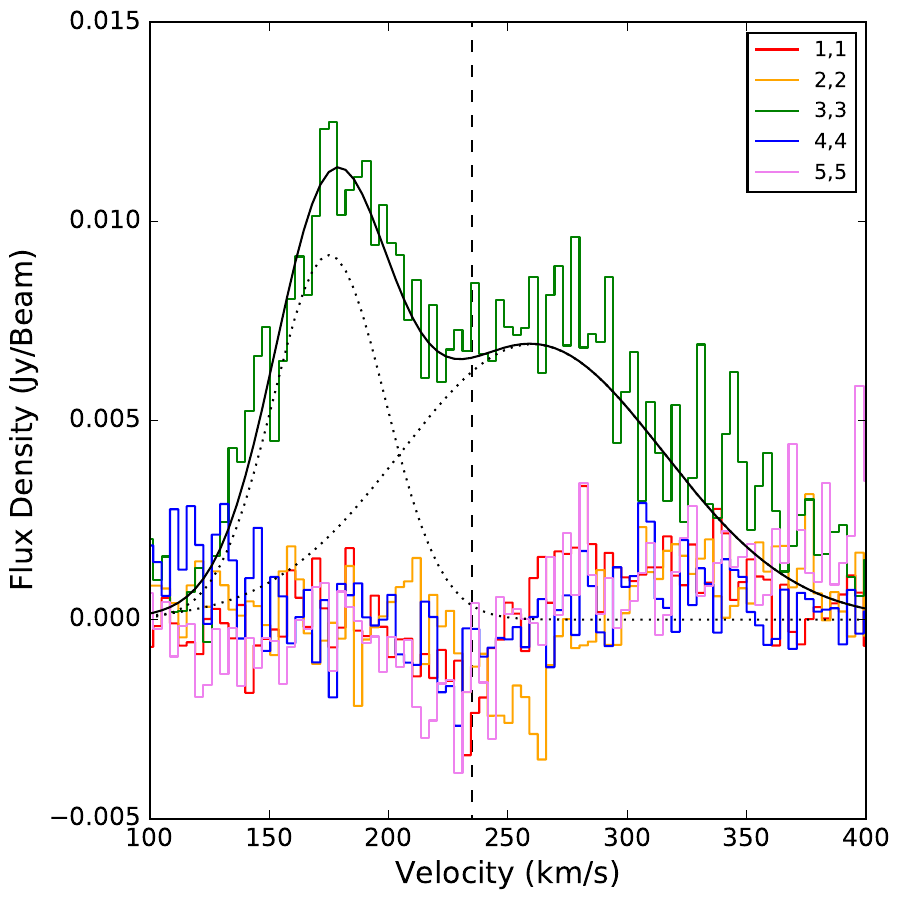}
\caption{The \amm\ spectra towards C1. The systemic velocity of 234 \kms\ is shown with a vertical dashed line. All the para species of \amm\ appear in absorption while the \amm(3,3) line appears in emission. The solid black line represents the best fit two Gaussian profile. The individual Gaussian components are plotted with dotted lines.}
\label{fig:33fit}
\end{figure}

A1, A5 and A6 FWHMs are a few tens of \kms\ wider than the other locations. These three sites are located on the edges of superbubbles (See section 4.4) discovered by \cite{Saka2006}. At C1, the \amm(3,3) line appears in emission whereas all other inversion lines appear in absorption (Figure \ref{fig:linemos}), interpreted to be due to the existence of \amm(3,3) masers, confirming the result from \citet{Ott2005}. The \amm(3,3) emission for C1 is not well described by a single Gaussian and thus two Gaussians were fit to the data (Figure \ref{fig:33fit}), with FWHMs of 55$\pm$3 \kms\  and 130$\pm$10 \kms\ and centers of 172$\pm$1 \kms\ and 257$\pm$5 \kms, respectively {\amm(3,3)a and \amm(3,3)b.}

\subsubsection{\water\, masers}

We identify three regions of \water\ maser emission in the data cube, labeled W1 to W3 in Figure \ref{fig:integrated} (center). The W1 water maser has been previously observed by \citet{Henkel2004} and \citet{Brun2009}. Spectra at these positions are shown in Figure \ref{fig:waterspec}. 
 Regions W1 and W3 are seen as clear, spatially resolved peaks in Figure \ref{fig:integrated}. Region W1 shows multiple velocity components with the main peak labeled W1a, and the minor peaks labeled W1b and W1c.
W2 is a faint feature which is not spatially resolved from W1 and W3, but marks the peak emission of a unique broad component shown by the shaded region in Figure \ref{fig:waterspec}. {W2 is identified by the peak emission from the narrow feature at 273 \kms\,channel} (Figure \ref{fig:waterspec}). W1 and W2 show multiple peaks in the spectrum. These were fitted with single Gaussians where appropriate. The properties are listed in Table \ref{tab:waterprop}. Single channel features are likely real, given our spectral resolution, but, were not fitted by Gaussians and thus their FWHMs are upper limits.

 The spectrum of W2 has contributions from W1 and W3  as they are not resolved from W2. W2 is a broad pedestal  spanning $\sim$100 \kms\ centered at 233 \kms\ with {several narrow features}. The rest of its properties are listed in Table \ref{tab:waterprop}.  This component is much better matched to systemic velocity of NGC 253. W3 is a redshifted single velocity component maser at 303 \kms. 

\begin{figure*}
\includegraphics[width=0.95\textwidth]{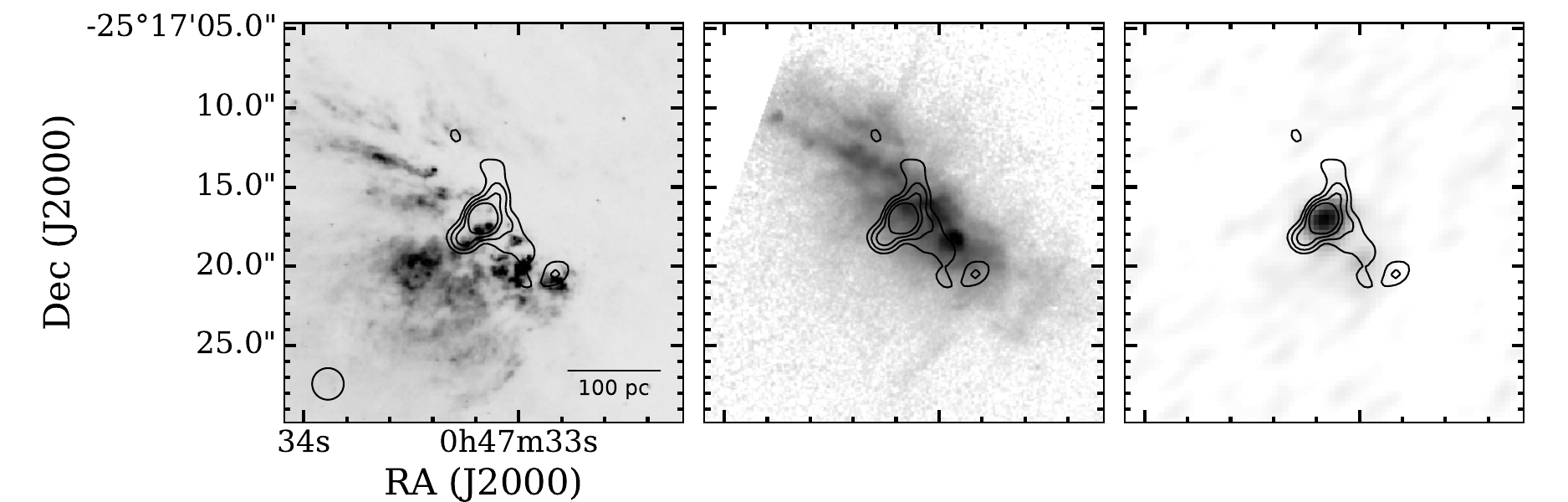}
\caption{Super-resolved \water\, 0.02, 0.04, 0.08, and 0.16 Jy beam$^{-1}$ \kms\ contours plotted on HST WFPC2 H$\alpha$ (left; \citealt{Watson1996}) map, HST P$\alpha$(center; \citealt{AH2003}), and RRL H56$\alpha$ (right). The synthesized beam in the bottom left corner of the left panel shows the resolution of the super-resolved cube.  }
\label{fig:superwater}
\end{figure*}

W1a dominates region W1 and is blueshifted with respect to systemic with an observed velocity of 109 \kms.   The integrated flux density of the W1a maser is $\sim 214$ K \kms\ yielding a luminosity of 0.66 \Lsun making it a kilomaser. 
The extension is hard to discern in the the peak flux density and integrated flux maps (Figure \ref{fig:integrated}), but it is clearly seen in the contours of the super-resolved image in Figure \ref{fig:superwater}, where contours of the super-resolved image cube are plotted on the HST H$\alpha$, P$\alpha$, and the RRL H56$\alpha$ images. The super-resolved  \water\ contours show the extension perpendicular to the major axis of NGC 253. This may indicate the W1 \water\ masers are related to the outflow of NGC 253 whereas W2 and W3 are spatially more consistent with the nuclear material. 
This result needs to be confirmed with higher resolution data. Lastly, we do not detect the 145 mJy \kms\ \water\ maser dubbed \water-2 from \cite{Henkel2004} observed in September 2002.  \water\ masers associated with YSOs and AGB stars can be variable on time scales of months (e.g. \citealt{Claussen1996}, \citealt{Felli2007}), therefore a non-detection is unsurprising.

\subsubsection{36 GHz \methanol\ masers}

Extragalactic 36 GHz \methanol\ masers were first detected by \citet{EC2014} in NGC 253 with the Australia Telescope Compact Array (ATCA)\footnote{The Australia Telescope Compact Array is part of the Australia Telescope, which is funded by the Commonwealth of Australia for operation as a National Facility managed by CSIRO.} with a 8.0\arcsec$\times$4.2\arcsec\ synthesized beam. 
Two sources were detected. The emission is likely not thermal in its nature, despite the large FWHM of the line, due to the total integrated intensity being $\sim$20 times greater than the total  integrated intensity from the Galaxy's central molecular zone (see \citealt{EC2014}). We resolve the two regions previously seen 
in \citealt{EC2014} into five different regions with masers
 as marked in the peak flux map.The spectra are extracted from the common resolution (6\arcsec$\times$4\arcsec) image cubes and shown in Figure \ref{fig:methanolspec}. The pairs M1 and M2, and  M4 and M5 are not spatially resolved from each other in the common resolution cube. M1 and M2 are spectrally similar, however M4 and M5 show distinct spectral components. The \methanol\ line is very close to the edge of one of our spectral sub-bands. There are small (3.5\kms) gaps between each sub-band where the data collected is untrustworthy, thus the channel corresponding to 160\kms\ is lost. The data in this channel and the two adjacent channels are thus unreliable. We next fit single component Gaussians where appropriate. The extracted properties of the lines are shown in Table \ref{tab:methanolprop}. The region M5 was fit with two Gaussians due to the double peak. The spectral features are narrower than the \amm\ (this paper) and \aco\ features \citep{BA2013} from the same locations, with measured FWHMs spanning a range of 30-80\kms. The narrower widths suggest that the 36 GHz \methanol\ emission is not tracing the entirety of the molecular gas.


\section{Discussion}
	

\subsection{\amm\ Temperatures}

One advantage of observing \amm\, is that many of its transitions are close in frequency space, and therefore can be observed with single telescope, a single observational setup, and under the same atmospheric conditions. 
{In addition, the $< 5$\% change in frequency between the transitions means that their respective uv coverage and the flux they resolve out are nearly identical.}
Rotation temperatures can then be derived from as many pairs of metastable (J = K) states as have been observed. Assuming optically thin conditions upper level column densities may be determined from:
\begin{equation}
N_u(J,K)=\frac{ 7.73\times10^{13}}{\nu} \frac{J(J+1)}{K^2} \int T_{mb}\,dv
\end{equation}
where $T_{mb}$ is the main beam brightness temperature, the column density of the upper inversion state, $N_u$, is in cm$^{-2}$, and the frequency ($\nu$) is in GHz. A rotation temperature (T$_{JJ'}$) is derived from a pair of the metastable states (J and J$^\prime$) by:
\begin{equation}
\frac{N_u(J^\prime,J^\prime)}{N_u(J,J)} = \frac{g^\prime_{op} (2J^\prime+1)} {g_{op} (2J+1)} \exp\Big( \frac{-\Delta E}{T_{JJ^\prime}}\Big)
\label{eq:ratio}
\end{equation}
where the difference in energy between states J and J$^\prime$, $\Delta E$, is in K (the corrected version of the equation in \citealt{HM2000} as shown in \citealt{Ott2005}) and the $g_{op}$ are statistical weights depending on the \amm\, species ($g_{op}$ = 1 for para-\amm\ and J$\neq3$n where n is an integer, and $g_{op}$ = 2 for ortho-\amm\  with J=3n).  Rotation temperatures derived from pairs of \amm\ transitions, for locations A1 to A7 are shown in Table \ref{tab:ammprop}. The rotation temperatures are best illustrated in the Boltzmann diagram shown in Figure \ref{fig:trot} (top). We have plotted weighted column densities on the vertical axis and energy above the ground state on the horizontal axis. The slopes between any two points are thus proportional to the inverse rotation temperature, i.e. cold gas shows steeper slopes than warm gas (Equation \ref{eq:ratio}). Notice at location C1 we see \amm\ in absorption against the continuum (Figure \ref{fig:linemos}) and thus the rotation temperature must be measured differently (see next paragraph). A3 is located $\sim0.5$ beams (53 pc) from C1, thus the emission line is likely partially absorbed, making rotation temperature measurements here unreliable. 
	
Measuring rotation temperatures from \amm\ absorption is not possible without knowing the excitation temperature $T_{ex}$ and the optical depth $\tau$. \cite{HM1995} describes the process of extracting total \amm\ column densities N(J,K) (the sum of both the upper and lower inversion states):
\begin{equation}
\frac{N(J,K)}{T_{ex}}=1.61\times10^{14}\frac{J(J+1)}{K^2\nu} \tau \Delta\nu_{1/2}
\end{equation}
and,
\begin{equation}
\tau=-\ln(1-\frac{|T_L|}{T_C})
\end{equation}
where the units of $\Delta\nu_{1/2}$(FWHM) are \kms, $T_L$ and $T_C$ are brightness temperatures of the line and continuum, respectively, and $\tau$ is the optical depth. We have no means to measure the transition-dependent excitation temperature, so for simplicity we assume that for each metastable transition $T_{ex}$ is equal. In this case the rotation temperature can still be derived from Equation \ref{eq:ratio} by substituting N(J,K) for N$_u$(J,K). Rotation temperatures for C1 are shown in Table \ref{tab:C1prop}. Since the column density at C1 depends on the excitation temperature, the values plotted for C1 in Figure \ref{fig:trot} are really N(J,K)/$T_{ex}$.  The spatial dependence of the rotation temperature is shown in Figure \ref{fig:trot} (bottom). 

\begin{figure}
\centering\includegraphics[width=0.5\textwidth]{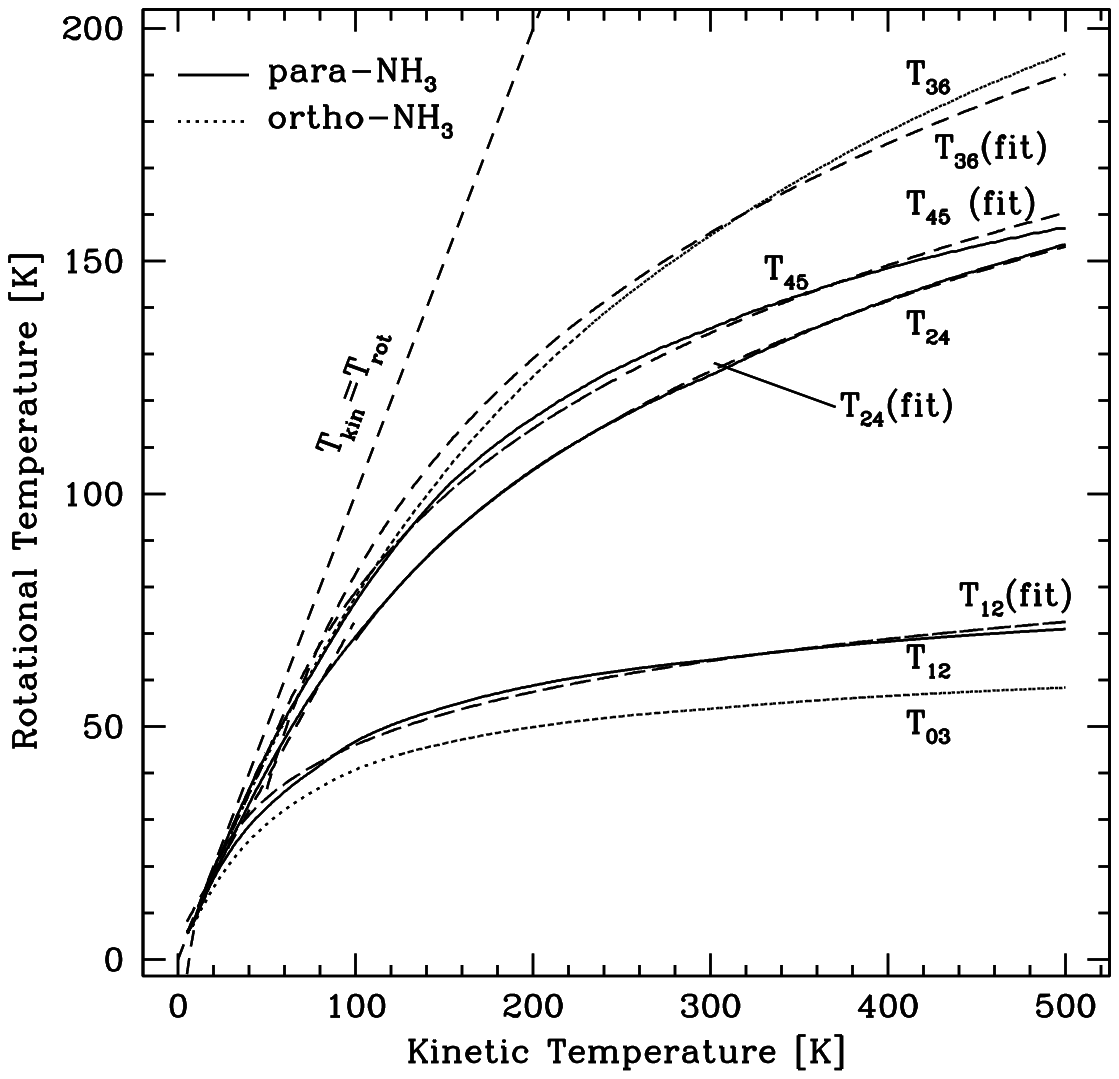}
\caption{The solid lines represent the relationship between the kinetic temperature and the rotation temperature. The dashed lines show the fits to the models represented by equations \ref{eq:24} and \ref{eq:45}, and equation 6 from \citet{OH2011}}
\label{fig:rot2kin}
\end{figure}

\begin{figure}
\includegraphics[width=0.5\textwidth]{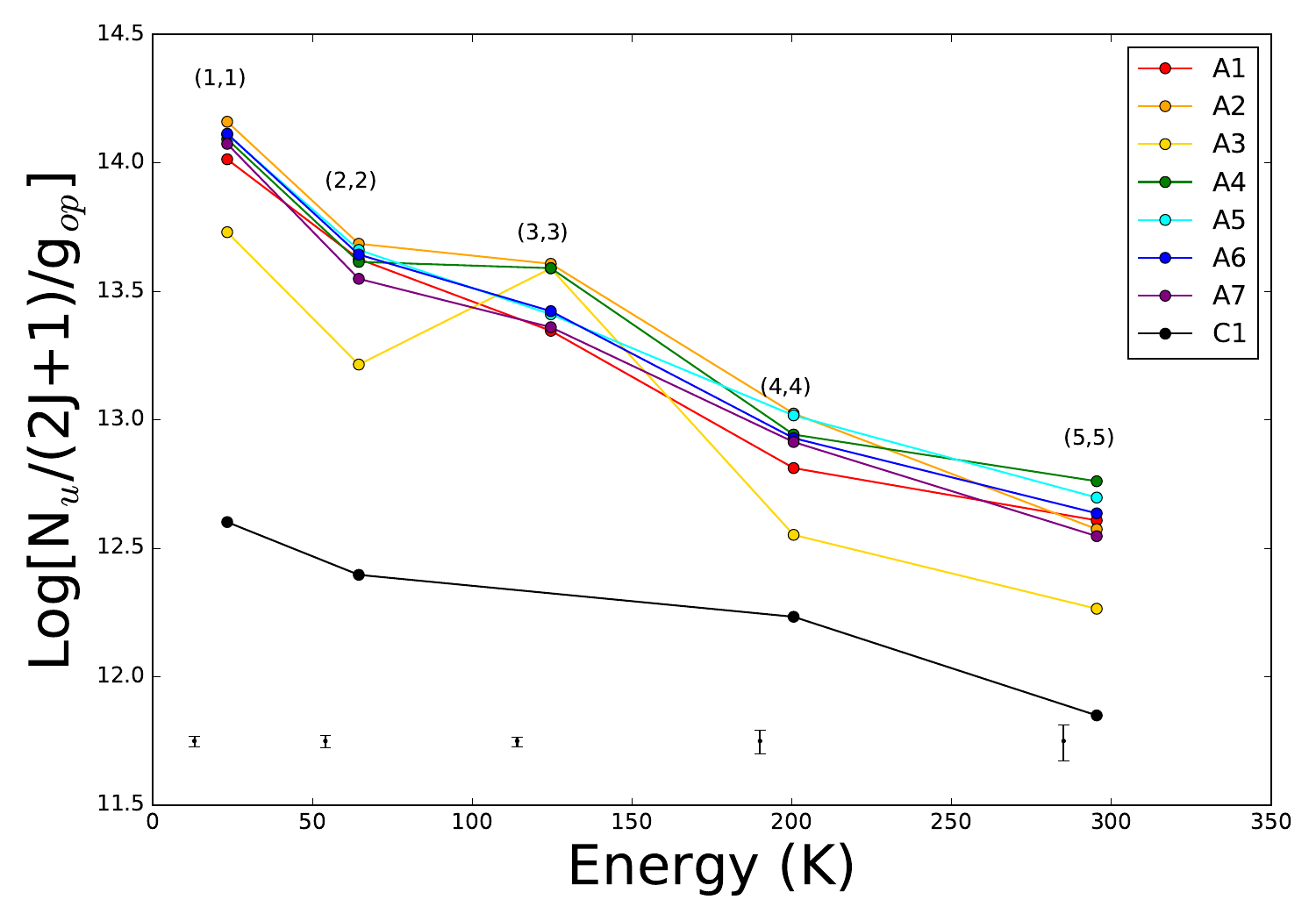}
\includegraphics[width=0.5\textwidth]{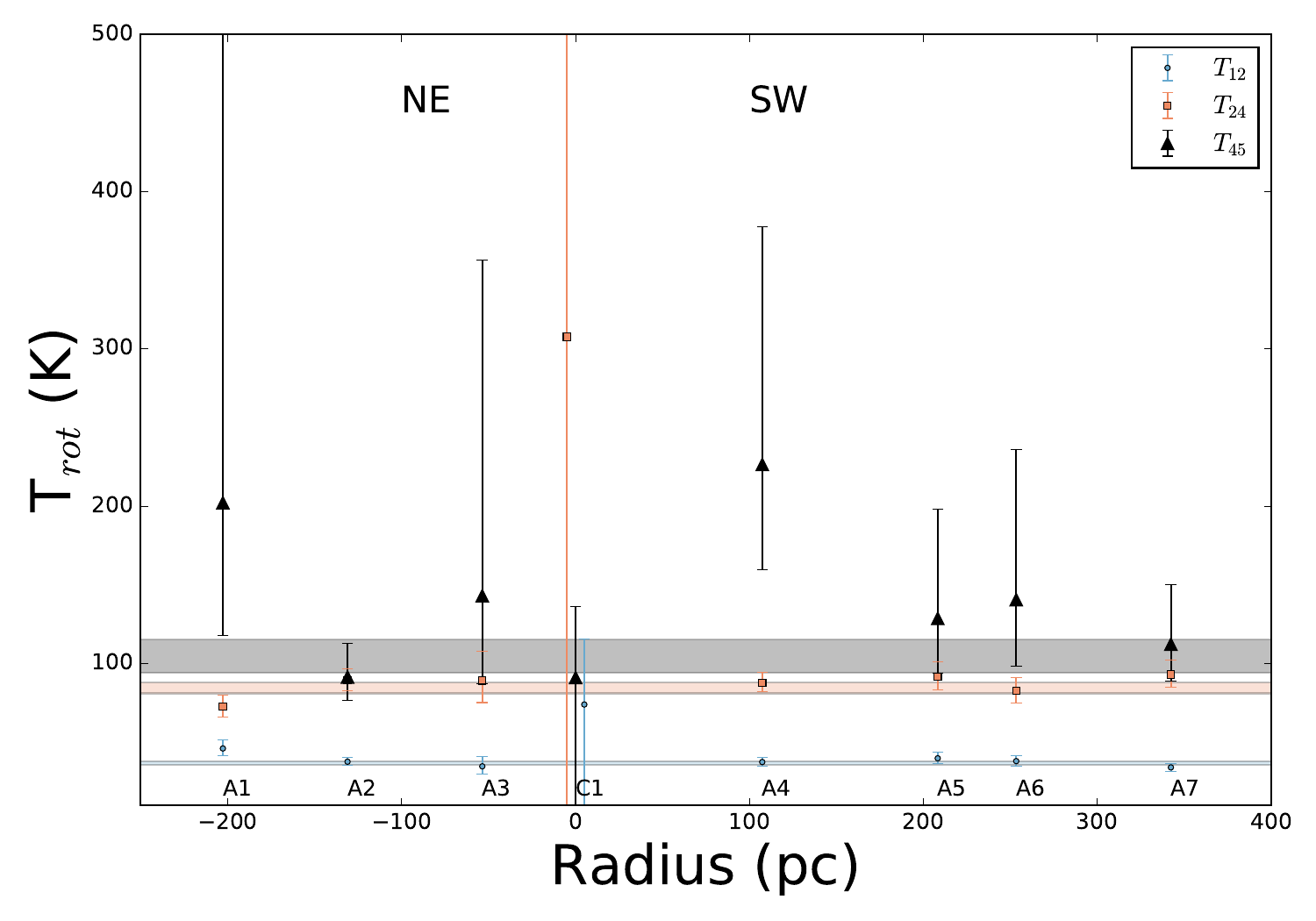}
\caption{Top: Boltzmann diagram of the observed \amm\ transitions. The black points representing C1 are the log of the normalized N/$T_{ex}$ not N$_u$, and have been scaled by a constant 0.85 preserving the proportionality of $T_{rot}$.  Error bars are plotted along the bottom and are the unweighted average of the individual metastable states from each location. Bottom: rotation temperatures derived from metastable line pairs. Error bars are measured from the minimum and maximum column densities of an individual line. The horizontal bars represent single temperature fits to the data. The height of the bar represents the uncertainty.  }
\label{fig:trot}
\end{figure}

The rotational temperature is not necessarily the true thermal temperature of the gas, i.e. it is not the kinetic temperature, but rather a lower limit. We therefore employ the functions from \cite{OH2011} to estimate kinetic temperatures. To convert rotation temperatures for the \amm(2,2) to (4,4) ratio additional fits to the same LVG models used in \citet{OH2011} from \citet{Ott2005} were made:
\begin{equation}
T_{Kin} = \begin{cases}
1.467 \times T_{24} - 6.984   &\text{for $T_{Kin}  \lesssim100K$}\\
27.085 \times \exp(0.019\,T_{24} ) &\text{for $T_{Kin} \gtrsim100K$}
\end{cases}
\label{eq:24}
\end{equation}
and for the \amm(4,4) to (5,5) ratio:
\begin{equation}
T_{Kin} = \begin{cases}
1.143 \times T_{45} - 1.611   &\text{for $T_{Kin} \lesssim50K$}\\
21.024 \times \exp( 0.0198\, T_{45}  ) ) &\text{for $T_{Kin} \gtrsim 50K$}
\end{cases}
\label{eq:45}
\end{equation}
Figure \ref{fig:rot2kin} shows how the conversion functions fit to the LVG models. We thus derive kinetic temperatures for adjacent pairs of like species for a total of 3 measurements at each location (Table \ref{tab:ammprop}). 

\begin{figure}
\centering
\includegraphics[width=0.5\textwidth]{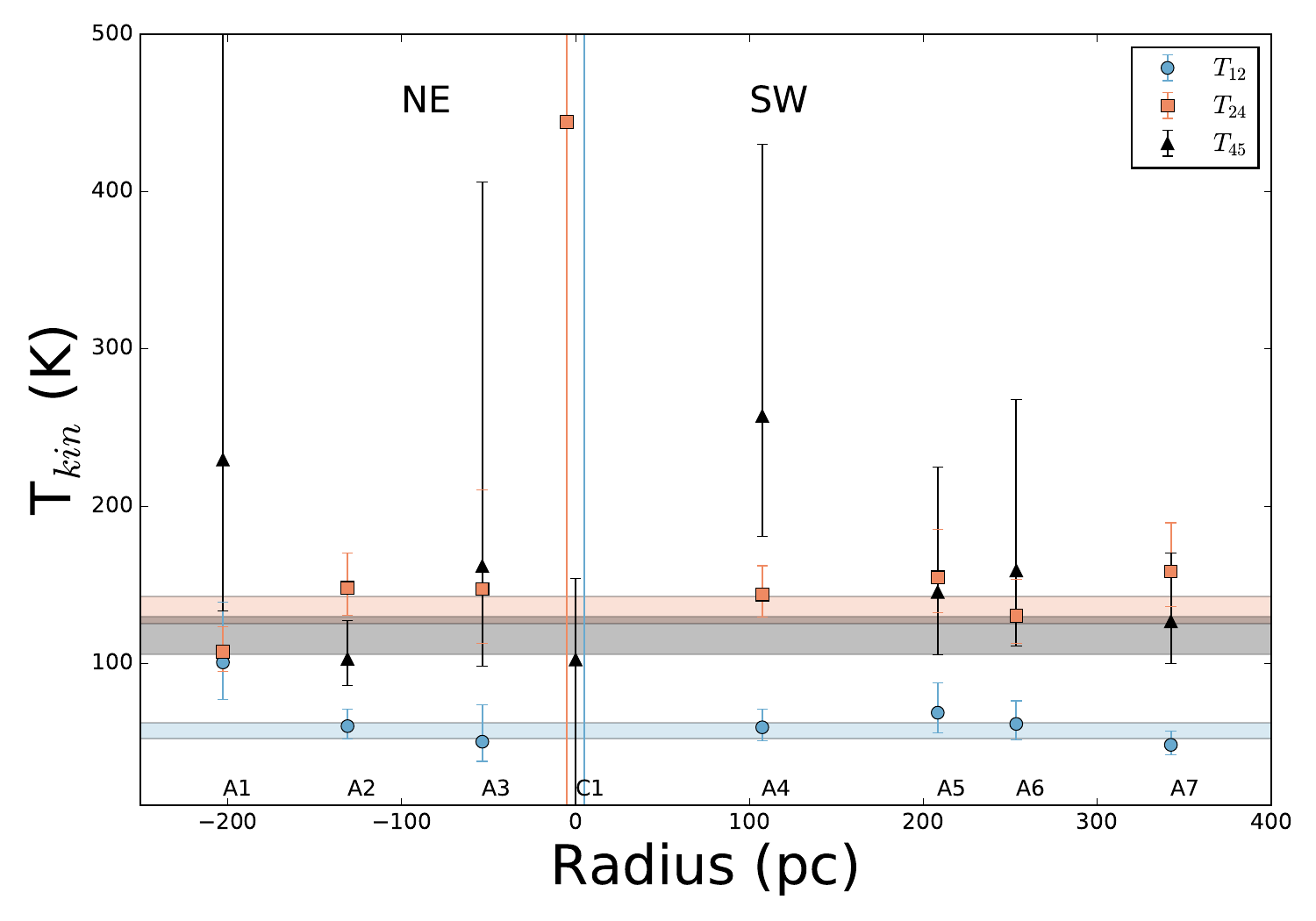}
\caption{Estimated kinetic temperatures after applying the LVG correction from \cite{Ott2005}. The horizontal bars represent single temperature fits to individual line pairs with the height representing the uncertainty. We see the need for a two temperature description of the molecular ISM. The derived temperatures from the J,K= 2,4,\& 4,5 ratios are consistent with a $\sim$130 K temperature component and the J,K=1,2 ratio measures a $\sim$57 K component. }
\label{fig:tkin}
\end{figure}

If the gas is dominated by a single kinetic temperature, all the rotation temperatures should yield that temperature. In the case that the rotation temperatures do not, either the gas must be represented by multiple kinetic temperatures, or the LVG approximation does not hold.  The LVG corrected kinetic temperatures are plotted in Figure \ref{fig:tkin}. The figures show remarkably little variance in the rotation and kinetic temperatures measured across all locations. We have fit a single temperature for each line pair across all locations weighted by the errors. We measure a weighted average for T$_{Kin 12}$ of $57\pm4$ K, for T$_{Kin 24}$ of $134\pm8$ K , and for T$_{Kin 45}$ of $117\pm16$ K. The T$_{Kin 24}$ and T$_{Kin 45}$ components are consistent within the errors. The kinetic temperatures of the dense molecular gas in the central kpc of NGC 253 is therefore most consistent with a cool $\sim$57 K derived from the (1,1) and (2,2) lines, and a warm $\sim$130 K component derived from the weighted average of the (2,2) to (4,4) and (4,4) to (5,5) ratios.

The \amm\ transitions of NGC 253 have been of interest to many others, but most recently have been observed with the Australia Telescope Compact Array (ATCA) \citep{Ott2005}, the VLA \citep{Takano2005}, and the Green Bank Telescope (GBT) \citep{Mangum2013}. Using interferometric observations of the bar, \citet{Ott2005} measure rotation temperatures T$_{12}\sim$42 K, and \citet{Takano2005}  measure T$_{12}\sim$26 K .  The lower temperature measured in \citet{Takano2005} is likely due to a low signal to noise ratio for the \amm(2,2) line.  Unlike what \citet{Ott2005} found, we see that a single temperature does not describe the dense molecular gas in NGC 253. \citet{Ott2005} observe the (1,1), (2,2), (3,3), and (6,6) lines,  while our analysis, that includes the \amm(4,4) and (5,5) lines, clearly indicates a warm component not observed by \citet{Takano2005} and a cooler component not observed by \citet{Ott2005}. The study by \citet{Mangum2013} indicates that there is a warm component to the dense molecular gas in NGC 253 as their analysis includes the \amm(4,4) line, however it is limited by the 30\arcsec\ beam of the GBT. This  limits their analysis to the NW and SE velocity components, measuring T$_{kin24}$ of 73$\pm$22 K and $<$150 K, respectively. It is possible that these measurements are affected by absorption in the center of NGC 253, or that they are sensitive to a more diffuse component of the molecular gas. On average they measure a kinetic temperature of 78$\pm$22 K from the \amm(J,K$\leq$4) transitions for the whole galaxy, which is broadly consistent with the results presented in this paper. Modeling of the CO ladder from J=4-3 to J=13-12 from \citet{Rosenberg2014} yields similar results. They find evidence for three temperature and density components for the molecular ISM ranging from 60 K to 110 K and 10$^{3.5}$ cm$^{-3}$ to 10$^{5.5}$ cm$^{-3}$, respectively. Their observations from the \textit{Herschel Space Observatory} have a resolution of 32.5\arcsec. Our \amm\ analysis does not reveal any information about the density of the molecular gas. Therefore our results are only broadly consistent in terms of temperature analysis. 

The spatially uniform distribution of temperatures is perhaps surprising due to the concentration of supernovae remnants in the central 500 pc \citep{UA1997}. We would expect that if heating by supernovae were a dominant effect in setting the state of a GMC, then an increase in temperature towards the center of the bar would be observed as the density of supernovae increases towards the center of NGC 253. This may not be true during the bulk of a GMC's lifetime: simulations by \citet{Murray2010} suggest that supernovae may only contribute to heating late in the GMC's life after much of the GMC has already been disrupted by jets and radiation pressure from stars. 
Heating may also not be observed because supernovae might {dissociate \amm\ molecules altogether in the region of greatest energy input.}

\subsection{LVG fitting with RADEX}	

To further investigate the need for multiple temperatures, we attempt to fit the data with Large Velocity Gradient (LVG) models directly. We do this to compare with the approximation to the LVG correction in section 4.1. We use RADEX \citep{RADEX} with collisional coefficients from the LAMBDA database \citep{Schoier2005}. RADEX was not used by \cite{Ott2005} and \cite{OH2011}, however it does make use of some of the same collisional coefficients. The collisional coefficients from the LAMBDA database cover temperatures up to 300K,  thus the kinetic temperature axis of the grid spans 0-300 K in steps of 3 K. The collider (H$_2$) volume density and \amm\ column density axes of the grid are logarithmically sampled respectively from $10^2$-$10^6$ cm$^{-3}$ , and from $10^{13}$ and $10^{17}$ cm$^{-2}$ with 100 steps each. { In the LVG approximation the ratio $N/\Delta v$ is the independent variable, however in RADEX the column density and line width are specified separately.} We specify a line width calculated from the weighted mean of  74\kms\ from all the para species of \amm\ in Table \ref{tab:ammprop}. 

\begin{figure*}
\centering
\includegraphics[width=1.0\textwidth]{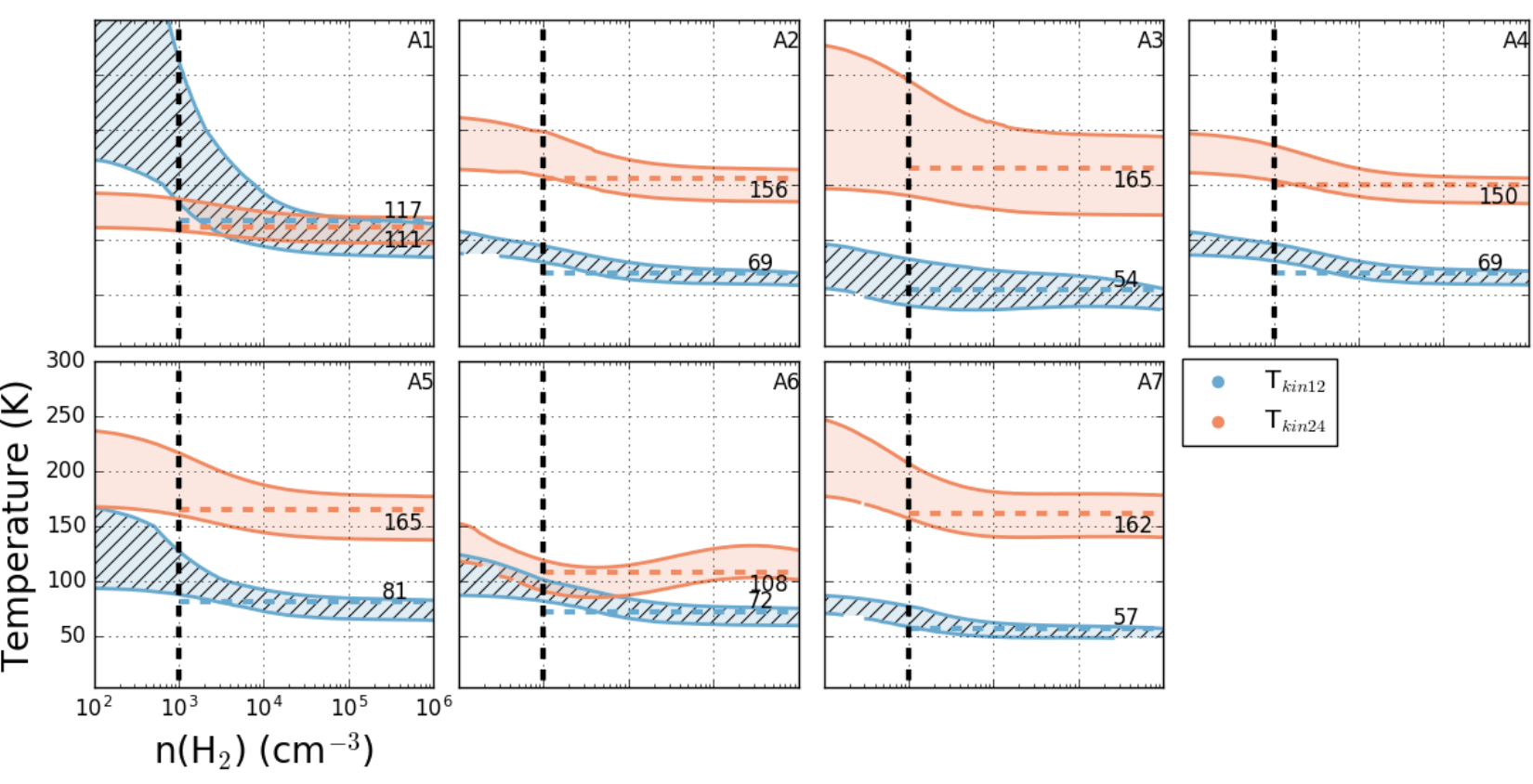}
\caption{Best fits of the LVG models to the data. The blue hatched region represents the \amm\ J=1,2 ratio, and the red region \amm\ J=2,4. The shaded areas represent the 1$\sigma$ errors in the fit. 
The horizontal dashed line represents the the median temperature fit for densities exceeding the critical density of \amm\ of 10$^{3}$ (vertical line)}
\label{fig:lvg}
\end{figure*}

The fits were carried out for each ratio of adjacent J=K para-\amm\ species. The results are tabulated in Table \ref{tab:LVG} and qualitatively displayed in Figure \ref{fig:lvg}. 
In the figure the median fitted temperature is drawn with a dashed line between the 1$\sigma$ confidence contours. The errors are calculated from the rms of the median fit for values above a critical density of $\sim10^3${ \cite{Cheung1968}. Below this density there is a slight upturn to the temperature parameter suggesting that high-T and low-n may excite ammonia at relatively low densities. This high-T low-n condition is unlikely as many higher critical density gas tracers have been observed with similar structure in the nucleus of NGC 253 \cite{Meier2015}.} The H$_2$ density is not well constrained, with error bars that span the entirety of the sampled parameter space, consistent with \amm\ not being a density probe. Generally, the fits are well behaved with the exception of the (4,4) to (5,5) ratio (not shown), for which the best fit solutions are poorly constrained and tend towards the edge of the temperature axis with values of $\sim$300 K. It is possible that there is a component that is hotter than our temperature range allows, but we cannot provide any meaningful estimates from the fitting. In contrast to the (4,4) to (5,5) ratio, the (1,1) to (2,2) and (2,2) to (4,4) ratios are well behaved. The solutions cover two regions of parameter space indicating a warm and cool component (Table \ref{tab:LVG}). The spatial distribution of temperatures is mostly uniform for the cool component, derived from line ratios \amm(1,1)/\amm(2,2),  with an average temperature of 
74$\pm$12 K, and  likewise for the warm component, derived from the \amm(2,2)/\amm(4,4) ratio,  with and average temperature of 145$\pm$14 K. This is broadly consistent with the analysis in \S4.1.

\subsection{Nature of the 36 GHz \methanol\ masers}

The nature of the 36 GHz methanol masers appears different from what we understand of their Galactic counterparts.  \cite{Yusef2013} performed a \methanol\ maser survey of the inner 160 x 43 pc of the Galaxy. We will use their study as a template to address \methanol\ emission in NGC 253. The \methanol\ line widths in NGC 253 span tens of \kms\, whereas individual Galactic \methanol\,masers tend to span a few \kms. The NGC 253 36 GHz line is spectrally resolved and multiple components can be fitted (see Table \ref{tab:methanolprop}). The maser with the largest flux found by \cite{Yusef2013} (number 164) is $\sim470$ Jy \kms, which corresponds to an isotropic luminosity, assuming a distance of 8 kpc, of $1.1\times10^{-3}$ \Lsun. By comparison the most luminous \methanol\ maser in NGC 253 is 1.63 \Lsun.   This would mean that the beam averaged class I \methanol\ masers in NGC 253 are about a thousand times more luminous than the most luminous \methanol\ maser found by \cite{Yusef2013}. Alternatively, there may be thousands of bright masers at similar velocities in one 6\arcsec$\times$4\arcsec\ beam.  
Surveys of Class I methanol masers in supernovae remnants (e.g \citealt{McEwen2014} \& \citealt{McEwen2016}) have revealed spectrally similar results to \cite{Yusef2013} with narrow spectral features spanning a few \kms.
At our spatial resolution of $101\times67$ pc the \methanol\ emission is not well resolved, and the entirety of the \cite{Yusef2013} survey corresponds approximately to the size of one of our beams.  Since there are $>300$ sources in a similar region of the Galaxy, and the NGC 253 spectra show multiple velocity components, it is possible that there are many components at each location.
The broad spectral profile could be constructed from many masers from protostellar outflows or supernovae remnants.
 
\subsection{ Impact of Superbubbles on the Dense Molecular ISM}
	
\begin{figure*}
\centering
\includegraphics[width=0.80\textwidth]{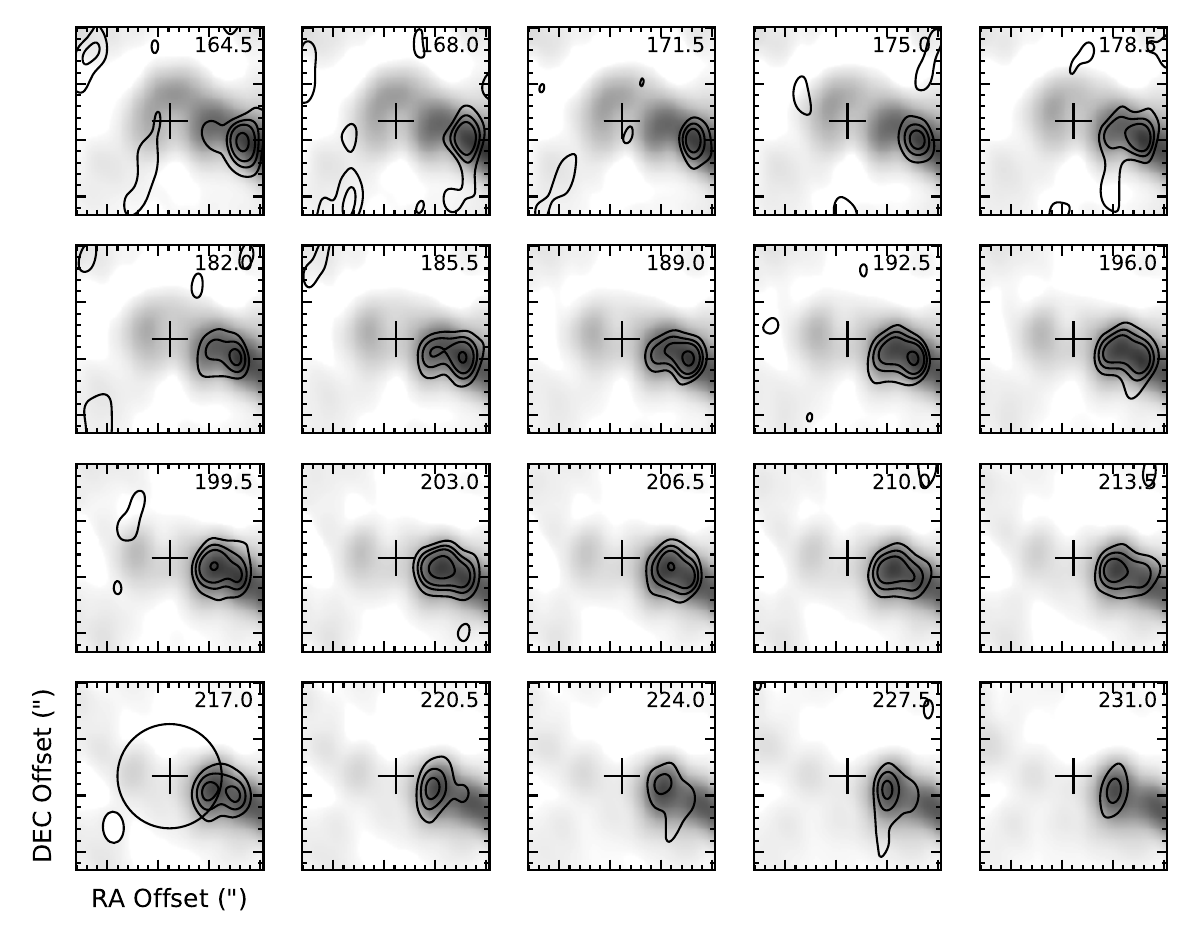}
\includegraphics[width=0.80\textwidth]{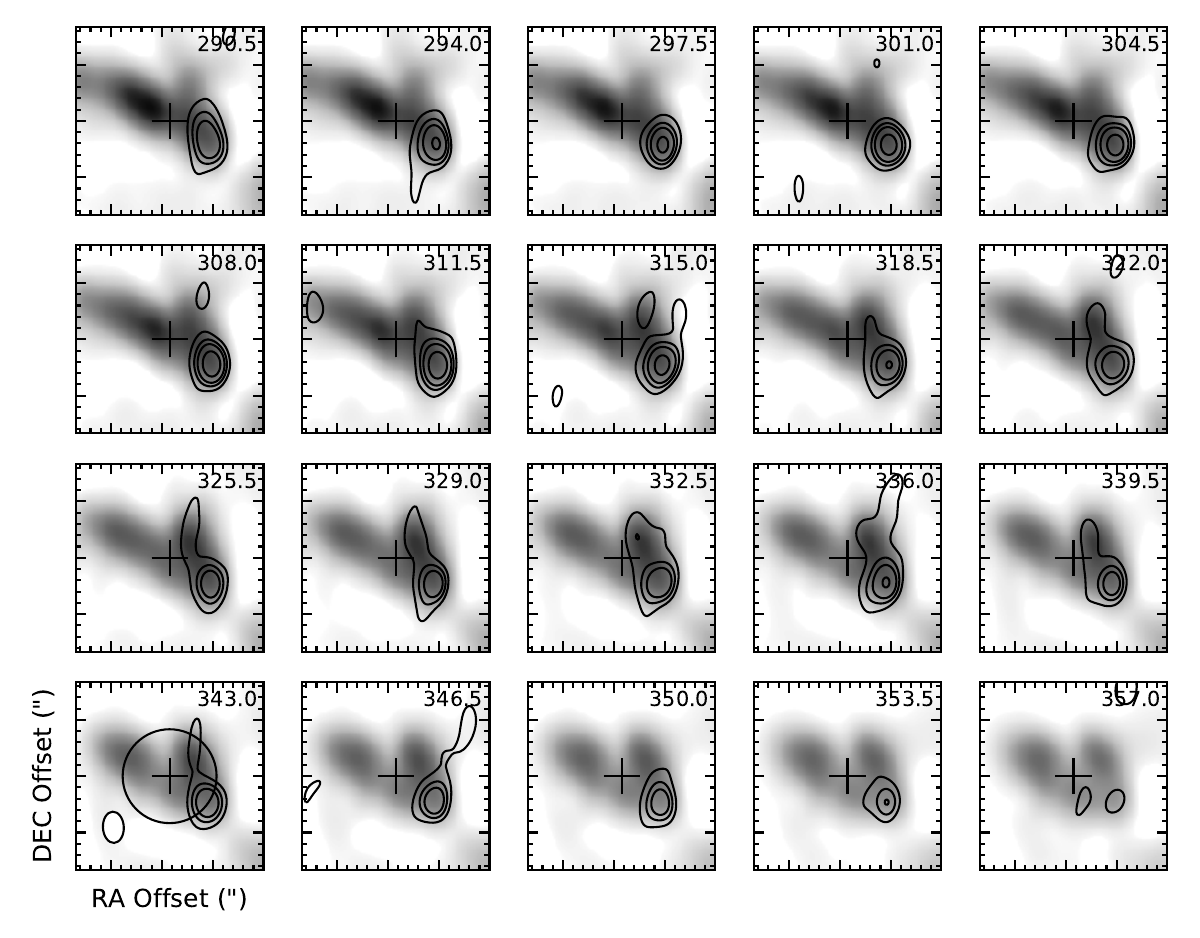}
\caption{Channel maps of the regions containing the superbubbles found by \cite{Saka2006}. The top block shows the eastern superbubble (RA: 00$^h$47$^m$34.64$^s$, DEC:  -25$^\circ$17'09.5") and the bottom block shows the western superbubble (RA: 00$^h$47$^m$32.53$^s$, DEC  -25$^\circ$17'24.0"). Each box is 18\arcsec$\times$18\arcsec. The black circle shows the diameter of the superbubbles measured by \citet{Saka2006}. The gray scale image shown is \aco\,emission from \cite{BA2013}. The 1, 2, 3, and 5 times 4.5 mJy beam$^{-1}$ contours of \methanol\ are shown and the velocity in \kms\ is shown in the top right of each panel.}
\label{fig:sprbbls}
\end{figure*}

\cite{Saka2006} find two superbubbles in NGC 253 in \bco\ emission. The shells are $\sim100$ pc in diameter, have masses of order $10^6$\Msun\, and kinetic energies of order $10^{46}$ J, suggesting winds and supernovae from a super star cluster, or a  hypernova, as the creation mechanism. Close to the location where these superbubbles interact with dense molecular gas, as traced by \amm, we observe \methanol\ masers. Figure \ref{fig:sprbbls} shows \aco\ channel maps of the superbubbles with \methanol\ contours overlaid. There are two groups of masers. The first  is on the northeast side of the galaxy and consists of M1, M2 and M3. The other group  consists of M4 and M5 on the southwest side (Figure \ref{fig:integrated}).  All the \methanol\ masers except M3 can be associated with the Sakamoto superbubbles. The association indicates that these masers exist where clouds are influenced by the expanding superbubbles. This relationship is also indicated by larger \amm\ linewidths at the locations A1, A5, and A6, which are nearest the masers and the superbubbles.

The larger observed linewidths are not likely a result of more turbulence within the clouds.  A comparison with the GMCs from \citet{leroy2015} suggests that the broader observed line widths in our data are likely a result of beam smearing effects. With a 2\arcsec\ beam \citet{leroy2015} are able to resolve individual clouds. Their clouds are clearly dominated by turbulence, as in all cases the thermal linewidth is $<1$ \kms\ as calculated from our \amm\ derived temperatures. Using an average temperature of 117 K derived from our \amm\ analysis, the mean thermal energy stored in the \citet{leroy2015} clouds is $\sim5\times10^{43}$J, whereas the mechanical energy stored, as derived from turbulent linewidths, is $\sim1\times10^{46}$J.  The turbulent linewidths of individual GMCs detected in \citet{leroy2015} appear unaffected by the impact of the expanding superbubbles, and in fact appear uniform across the entire central molecular bar (Figure \ref{fig:bardisp}).  The larger \amm\ linewidths observed with a larger beam suggest instead that the superbubbles are imparting mechanical energy resulting in bulk translational motion of the GMCs. 

\begin{figure}
\centering
\includegraphics[width=0.5\textwidth]{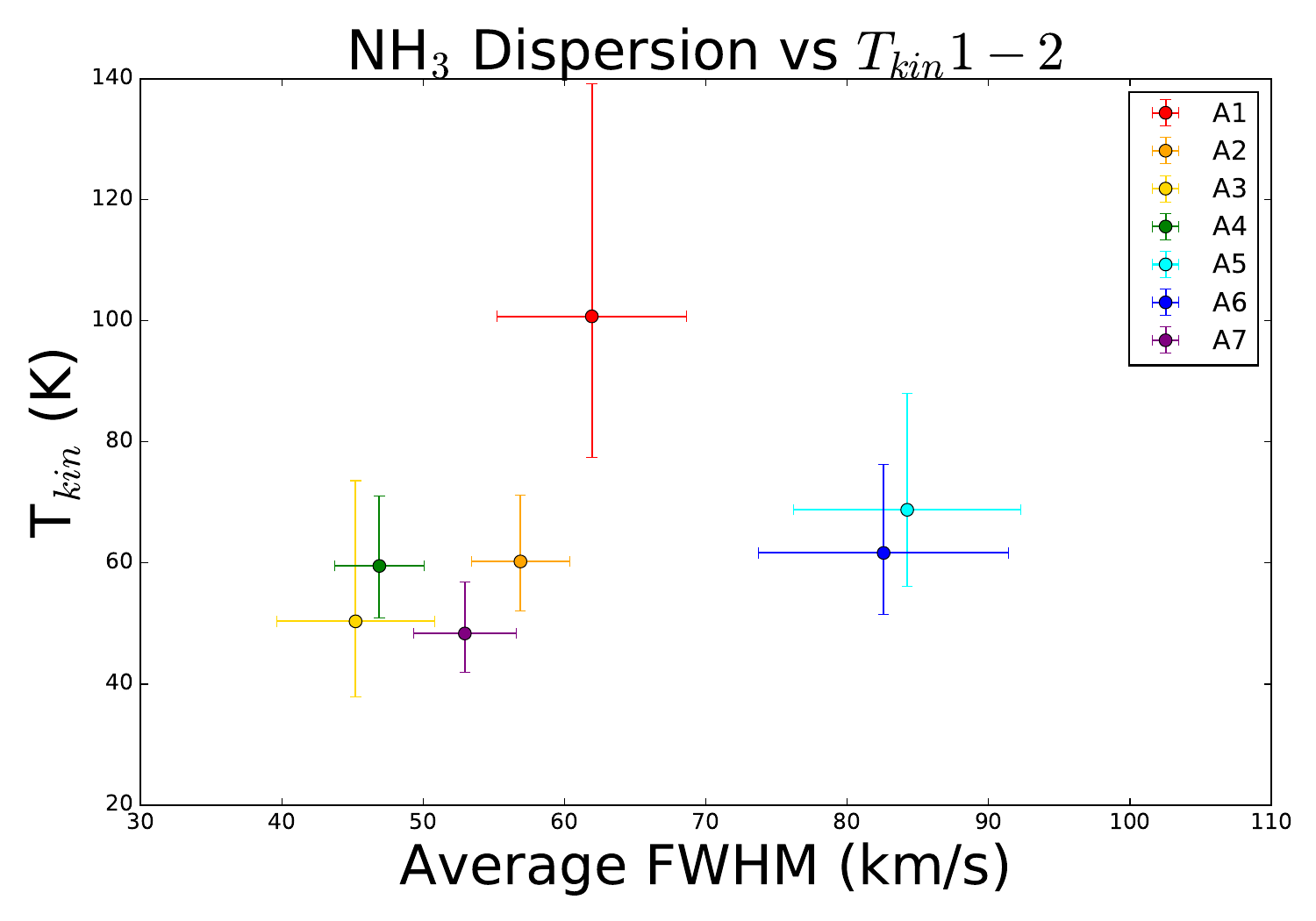}
\includegraphics[width=0.5\textwidth]{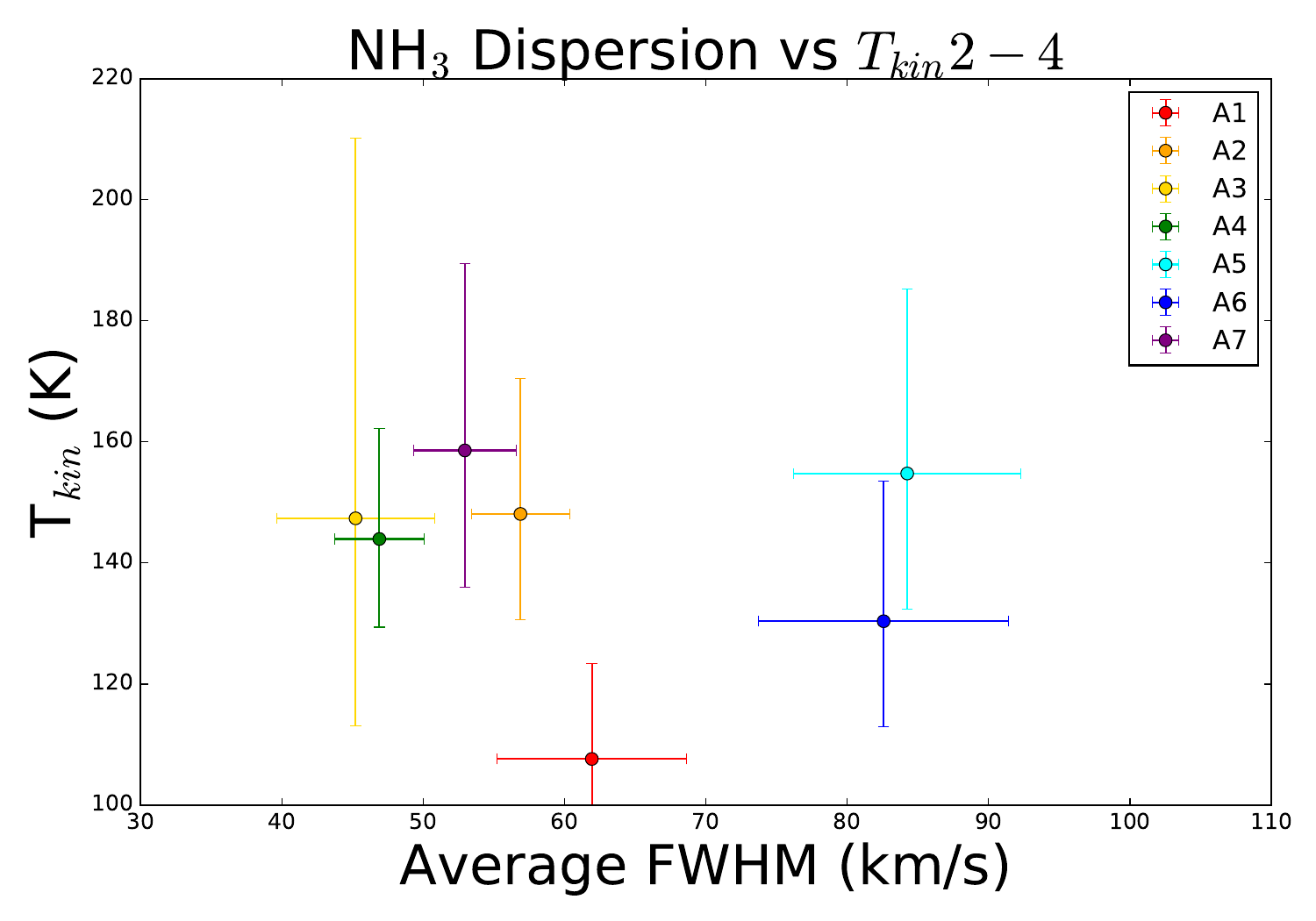}
\caption{Plots of the average FWHM vs. derived kinetic temperatures from pairs of the \amm\ J=1,2 (top) and J=2,4 (bottom) lines. }
\label{fig:disps}
\end{figure}

\begin{figure}
\centering
\includegraphics[width=0.49\textwidth]{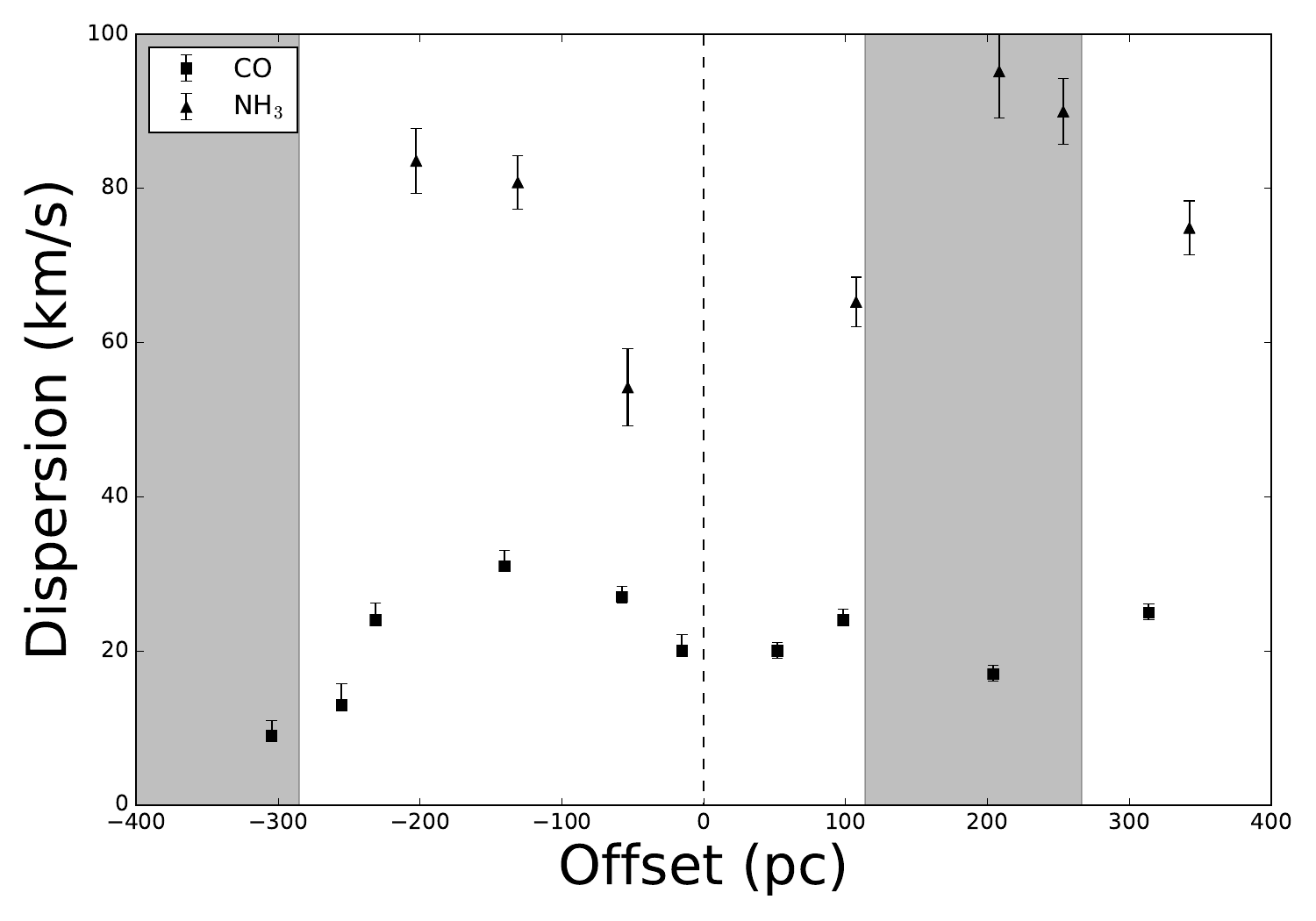}
\caption{The measured FWHM as a function of offset from the continuum peak C1. The circles represent the FWHM of \aco\ from \cite{leroy2015} and the triangles represent the \amm(1,1) FWHM. The areas of the expanding superbubbles are shaded grey.}
\label{fig:bardisp}
\end{figure}

The relationship of the clouds to the masers is shown by Figure \ref{fig:disps}, which plots  kinetic temperatures T$_{kin12}$ and T$_{kin24}$ derived from \amm\ against the mean FWHM of the \amm\ transitions. The figure shows little temperature variation toward the shocks traced by \methanol. If there is shock heating of the dense molecular gas it must be highly localized to areas much smaller than the beam such that it does not affect the overall kinetic temperature measured with $\sim$100 pc spatial resolution.

 \subsection{The Outflow, Masers, and Starburst}
 

 \subsubsection{\amm(3,3) masers}
 
The location of the continuum peak (C1) is also the locus of the starburst and the central base of the molecular and ionized bipolar outflow. Here the para-\amm\ lines are not observed in emission, however the ortho-\amm(3,3) line is. We did not detect any other ortho species of \amm\  at this location to compare with, however \cite{Ott2005} did observe the ortho (6,6) species to be in absorption at our location C1.  Since \amm(3,3) is the only ortho-\amm\ species observed in emission at this location, we corroborate the interpretation from \cite{Ott2005} that the \amm(3,3) line is masing here. The spectral line (Figure \ref{fig:linemos}) could not be fit with a single Gaussian, but rather two Gaussians  centered at 172\kms\ and 257\kms\ with widths of {55\kms\ and 130\kms,} respectively \amm(3,3)a and \amm(3,3)b (see Figure \ref{fig:33fit}).    The existence of the \amm(3,3) maser suggests that the collision rate in the center of NGC 253 increases or there exists an excess of infrared photons in the center in order to pump this maser. Currently, there is one other known extragalactic source of \amm(3,3) masers apart from NGC 253:  the Seyfert galaxy NGC 3079 \citep{Miyamoto2015}. Both galaxies host outflows, but NGC 3079 is host to an active galactic nucleus (AGN). It hosts a star formation rate of 2.6 \sfr\ over 4 kpc \citep{Yamagishi2010} measured from 1-1000$\mu$m emission. {Attribution to} an AGN driven wind might be favored because the starburst is weak. Since the mechanism driving the outflows in these galaxies is different, we hypothesize that in both cases it is the collision of the hot ionized outflow cone with the surrounding material that results in the \amm(3,3) masers.

\subsubsection{\water\ masers}
	
The \water\ masers are located within the centermost 200 pc. Being at the center of the starburst it is likely that all these masers are star formation related. \cite{Brun2009} investigated this location with the VLBA and inferred a pure starburst nature of the masers due to their similarity to Galactic \water\ masers and spatial coincidence with supernovae remnants. Our data, while much lower in resolution, indicate that W1 may be extended perpendicular to the disk and aligned with the bipolar ionized gas outflow (Figure \ref{fig:superwater}). The spectrum shows evidence of multiple components (Figure \ref{fig:waterspec}). W2 contains many individual components within a velocity space $\sim$100\kms\ wide centered at systemic velocity and includes contributions from W1 and W3. W3 is a single component maser located to the southwest of the nucleus. 

W3 is the simplest \water\ maser to explain. It is cospatial with an \hii\ region seen in Figure \ref{fig:SFmap} and has a narrow spectrum. It is likely associated with a massive star or stars in that region. Its isotropic luminosity of $2\times10^{-2}$ \Lsun\ is consistent with luminosities of AGB and YSO \water\ masers in the Galaxy (e.g \citealt{Palagi1993}). 

The \water\ maser W1 is the clear oddity in NGC 253. W1 is most clearly seen in the super-resolved image Figure \ref{fig:superwater}. There are three components with observed velocities in the range 25-109 \kms, while the systemic velocity of the galaxy is 234 \kms. The W1 maser emission does not follow the kinematics of the dense molecular ISM (Figure \ref{fig:waterspec}). The most luminous and broad component is centered at 109\kms.  According to \cite{Brun2009} this maser emission is associated with supernova remnant TH4 (\citealt{TH1985} and \citealt{UA1997}). However, because of the extreme conditions under which \water\ is known to mase, in addition to the locations and velocities, and because no \water\ masers are found in Milky Way supernova remnants (\citealt{Claussen1999} and \citealt{Caswell2011}), we hypothesize that these masers are tracing shocks in entrained material in the outflow. 

We consider three possibilities for the origin of the masers. First, Figure 11 in \cite{Strick2002} show possible anatomies of starburst driven outflows in NGC 253.  In this picture the W1 \water\ masers likely trace dense gas clumps closest to the starburst center or shocked dense gas entrained in the outflow. The observed velocities are consistent with the Strickland model predictions for outflow velocities of order 100\kms. A model of a conical outflow developed by analyzing optical integral field unit data in \cite{Westm2011} presents a more direct view of the outflow in NGC 253. Their data show that the ionized gas associated with the base of the outflow is blueshifted $\sim$100$\pm$50\kms\ with respect to systemic. This is a good match to the observed velocities of W1a to W1c suggesting that there are shocks related to the outflow. The origin of the shocks remains unknown as we do not see a spatial separation of W1a, b and c, even in the super-resolved image cube. We hypothesize that they may be shocks in entrained material in the outflow, or collisions with dense gas that funnels the outflow out of the galactic plane. In this picture W2 does not have observed velocities that match either the \citet{Strick2002} or \citet{Westm2011} models of the outflow. The velocity center of W2 matches the systemic velocity of NGC 253, suggesting that it may exist in a molecular torus about the center. The velocities of the \amm(3,3) masers and the \aco\ spectrum from the same location are similar (Figure \ref{fig:waterspec}). In the Strickland models the \amm(3,3) and W2 \water\ masers would exist where the outflow impacts the unperturbed disk triggering star formation. The W2 \water\ narrow components are likely generated by YSOs in the nuclear starburst, and the \amm\ masers are likely star forming sites like that of DR 21 \citep{Wilson1990} or W51 \citep{Goddi2015} in the Galaxy.

The second possibility is that, while the W1 masers still originate in the outflow, the W2 masers do not trace a molecular torus, but rather their velocity extent is due to a two-sided outflow. Here the portion of W2 redshifted with respect to systemic is tracing the receding side of the outflow. However, the receding side is not as well understood because it is obscured by dust and tipped away from the observer, making optical and soft X-ray observations difficult. \cite{Westm2011} find it impossible to model the receding outflow cone, and  the analysis from \cite{Strick2002} is focused on kpc scales. Therefore current evidence for the receding outflow is limited to hundreds of pc displacement from the disk. As \water\ masers are unaffected by dust obscuration, this picture implies fewer shocked regions as traced by \water\ on the receding side of the outflow, or that other masers in the receding outflow are unobserved due to line of sight effects. In this case the W1 masers would exist as in the first picture. 

Lastly, it is possible that the W1 \water\ masers are not related to the outflow at all, leaving us with the difficulty of explaining their velocities. \citet{Brun2009} noticed that W1 is cospatial with the supernova remnant TH4. The W2 masers in this case would likely be gas in the molecular torus as in the first picture or associated with known supernova remnants at their location. With an rms of 12 mJy it is unlikely that \citet{Brun2009} would have seen the fainter masers comprising W2. Our spatial resolution makes this problem difficult to solve, thus deeper high-resolution observations are necessary.


\subsection{Millimeter Molecular Lines}

\begin{figure}
\centering
\includegraphics[width=0.5\textwidth]{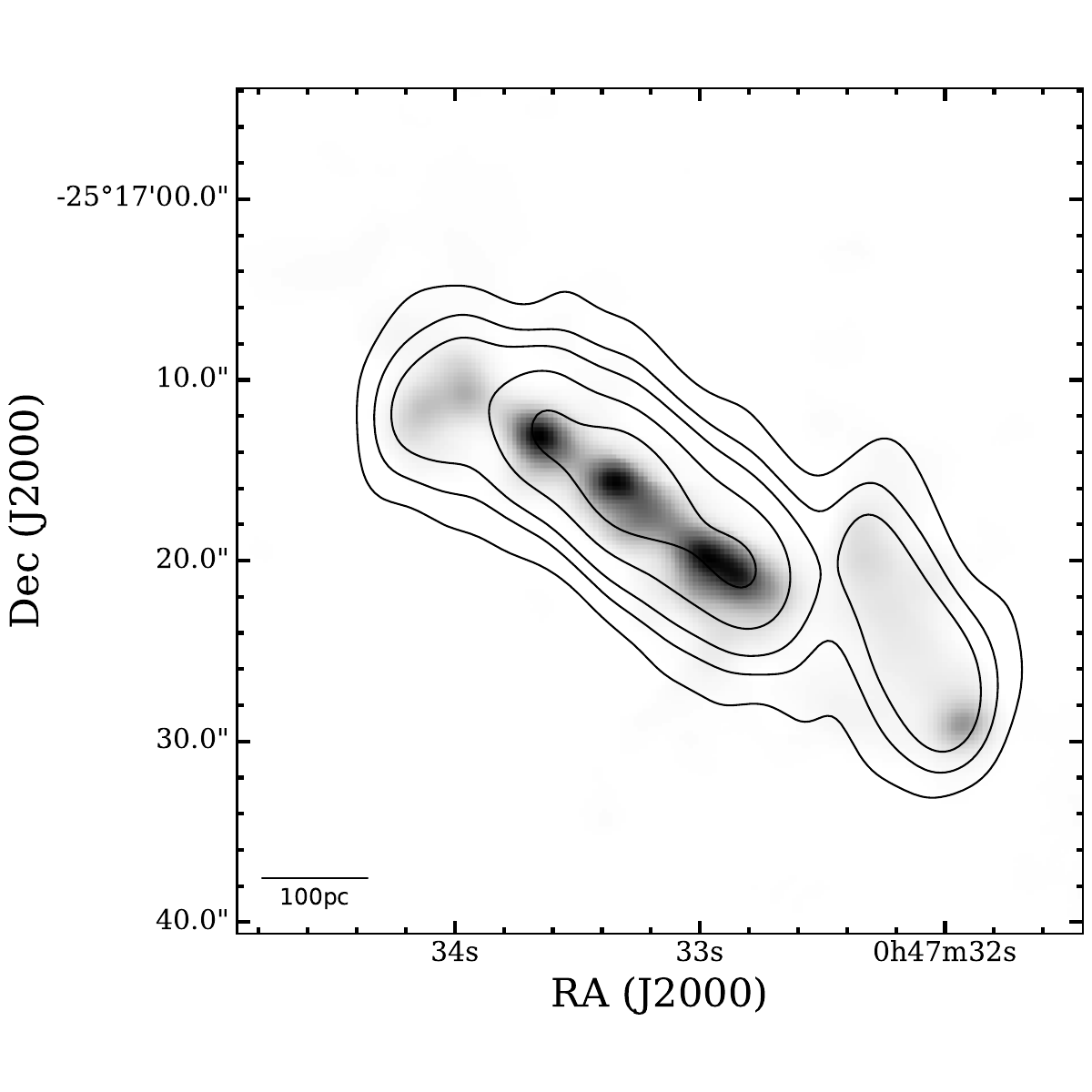}
\caption{HCN map from \cite{Meier2015} in greyscale with \amm(3,3) 3, 6, 9 ,15, and 24 $\times$ 4.8 K \kms\ contours.}
\label{fig:HCN}
\end{figure}

\begin{figure*}
\centering
\includegraphics[width=1.0\textwidth]{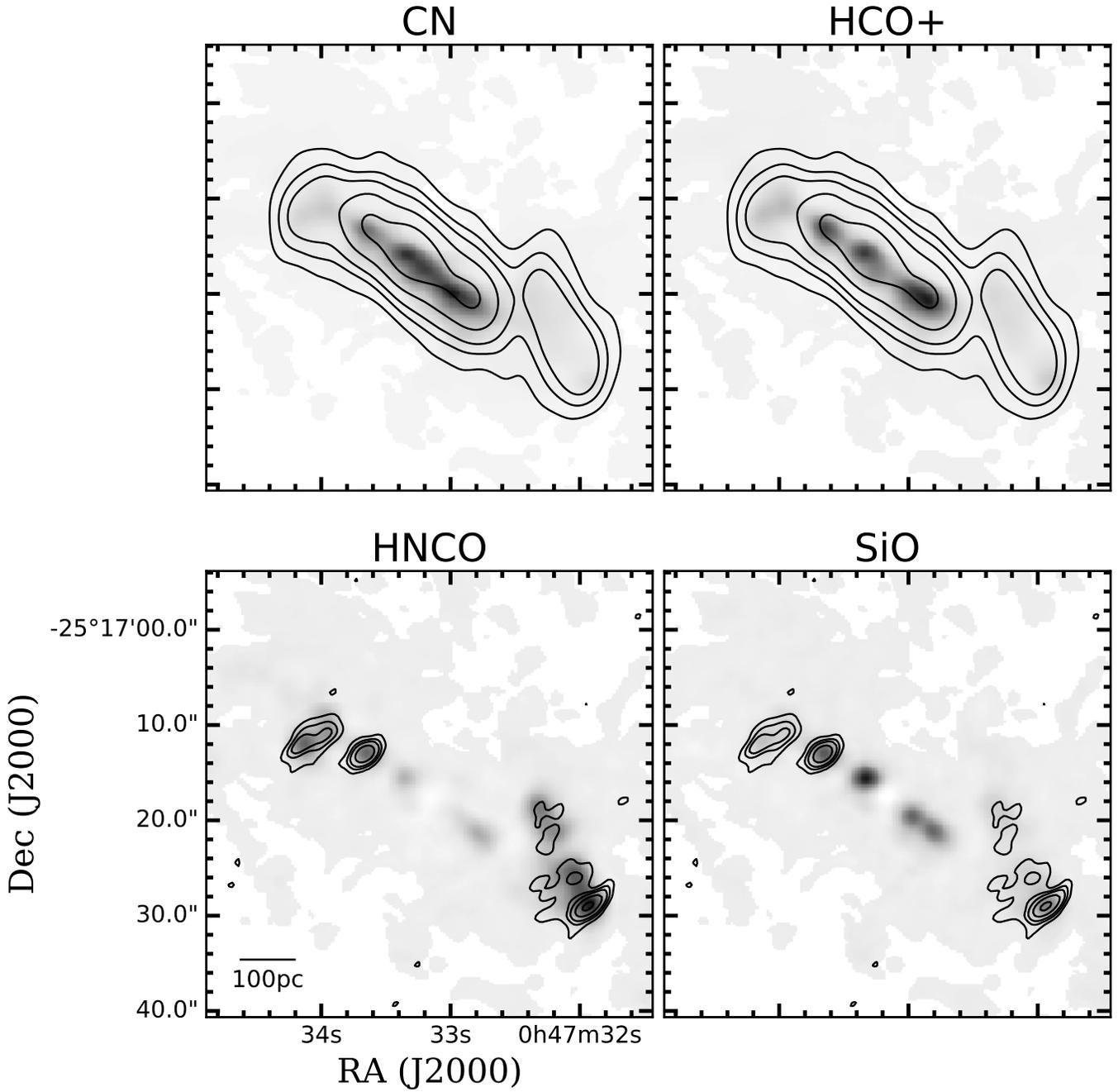}
\caption{The top row shows the PDR tracer CN(left) and intermediate density tracer HCO$^+$(right) \citep{Meier2015} with 3, 6, 9 15, and 24 $\sigma$ contours of \amm. The bottom row shows shock tracers HNCO and SiO \citep{Meier2015} with 3, 6, 9 15, and 24 $\sigma$ contours of \methanol. }
\label{fig:chem}
\end{figure*}

In order to reveal the underlying properties responsible for the observed temperatures and masers,  we compare with images of several molecular species from \cite{Meier2015} who observed NGC 253 with ALMA in the millimeter range.We have chosen five molecules that trace PDRs (113.40 GHz \cn ; \citealt{RodriguezFranco1998}), intermediate densities (89.18 GHz HCO$^+$ (1-0))
found in both diffuse and dense clouds \citep{Turner1995}, gas densities greater than $3\times10^4$ cm$^{-3}$ (88.63 GHz HCN (1-0) \citealt{Gao2004}), weak shocks (89.92 GHz HNCO ($4_{0,4}-3_{0,3}$ \citealt{MT2005}), and strong shocks (86.84 GHz SiO (2-1; $\nu=0$) \citealt{GB2000}). {The largest spatial scale of in the sampled by \citet{Meier2015} is ~18\arcsec\.They compare the ALMA interferometric maps with single dish observations and find that 50\%-100\% of the total flux is recovered.  The largest spatial scale of the VLA D configuration at K band of 66\arcsec\, thus we expect less flux to be resolved out by the VLA than for ALMA. The maximum uncertainty of the VLA to ALMA line ratios could consequently be increased by a factor of two.} The resolution of the ALMA cubes is $\sim$2\arcsec$\times$2\arcsec.  The HCN map, with \amm(3,3) contours in white, is shown in Figure \ref{fig:HCN}. Both molecular morphologies are well correlated. This suggests that these molecules are tracing similar environments  We show the PDR and dense gas tracers, CN and HCO$^+$, with \amm(3,3) contours overlaid (Figure \ref{fig:chem} top).Like HCN, CN and HCO$^+$ trace similar regions as \amm.  

\begin{figure*}
\centering
\includegraphics[width=1.0\textwidth]{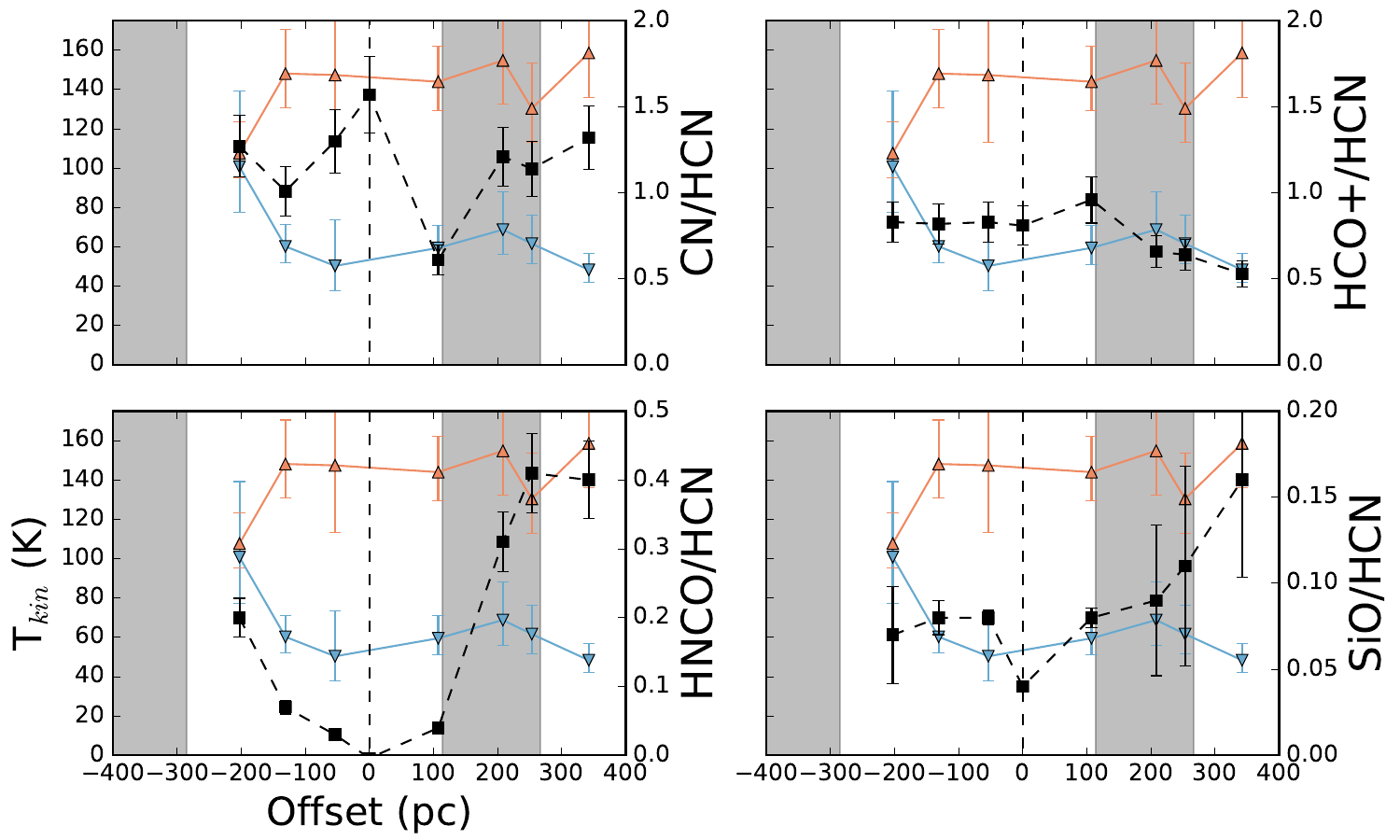}
\caption{ Intensities of molecular tracers relative to  HCN from the A1-A7 and C1 locations (dashed lines with black squares). The top row shows the PDR tracer CN (left) and intermediate density tracer HCO$^+$ (right) and the bottom row shows weak shock  tracer HNCO (left) and strong shock tracer SiO (right). The red triangles and blue inverted triangles with lines are the warm and cool components of the kinetic temperatures we derive from \amm, respectively. The grey bars represent the diameter and locations of the superbubbles in \citet{Saka2006}.}
\label{fig:chemratios}
\end{figure*}

For the shock tracers, HNCO and SiO, we have overplotted the naturally-weighted unsmoothed 36 GHz \methanol\ contours (Figure \ref{fig:chem} bottom). The \methanol\ masers are well correlated with HNCO and SiO at the sites of the \citet{Saka2006} superbubbles, however the centermost 200pc remains bright in SiO whereas HNCO dims \citep{Meier2015}. The \methanol\ masers appear much better correlated with HNCO than SiO. We take the dominant source of error in the individual molecular maps to be the 10\% estimate in the absolute flux density calibration. To estimate relative intensities the individual ALMA maps were smoothed to the common resolution of 6\arcsec$\times$4\arcsec\ to match our VLA data. We then normalized these maps by the HCN map. Line ratios were then measured at locations A1-A7 and C1. The line ratios are shown in Figure \ref{fig:chemratios} plotted in black with the T$_{Kin12}$ shown in red and T$_{Kin24}$ shown in blue.

The cold and warm temperature components of the molecular gas do not correlate well with PDR tracers (CN), intermediate densities (HCO$^+$), weak shocks (HNCO), or strong shocks (SiO). We interpret the plots in Figure \ref{fig:chemratios} to mean that none of the selected processes relate to the heating and cooling of the dense molecular ISM on 100 pc scales in NGC 253. This possibly means that cosmic rays are the dominant source of heating in the center of NGC253, {or that the dominant heating source destroys} \amm\ molecules and we are only sensitive to less impacted molecular gas

\begin{figure*}
\centering
\includegraphics[width=0.95\textwidth]{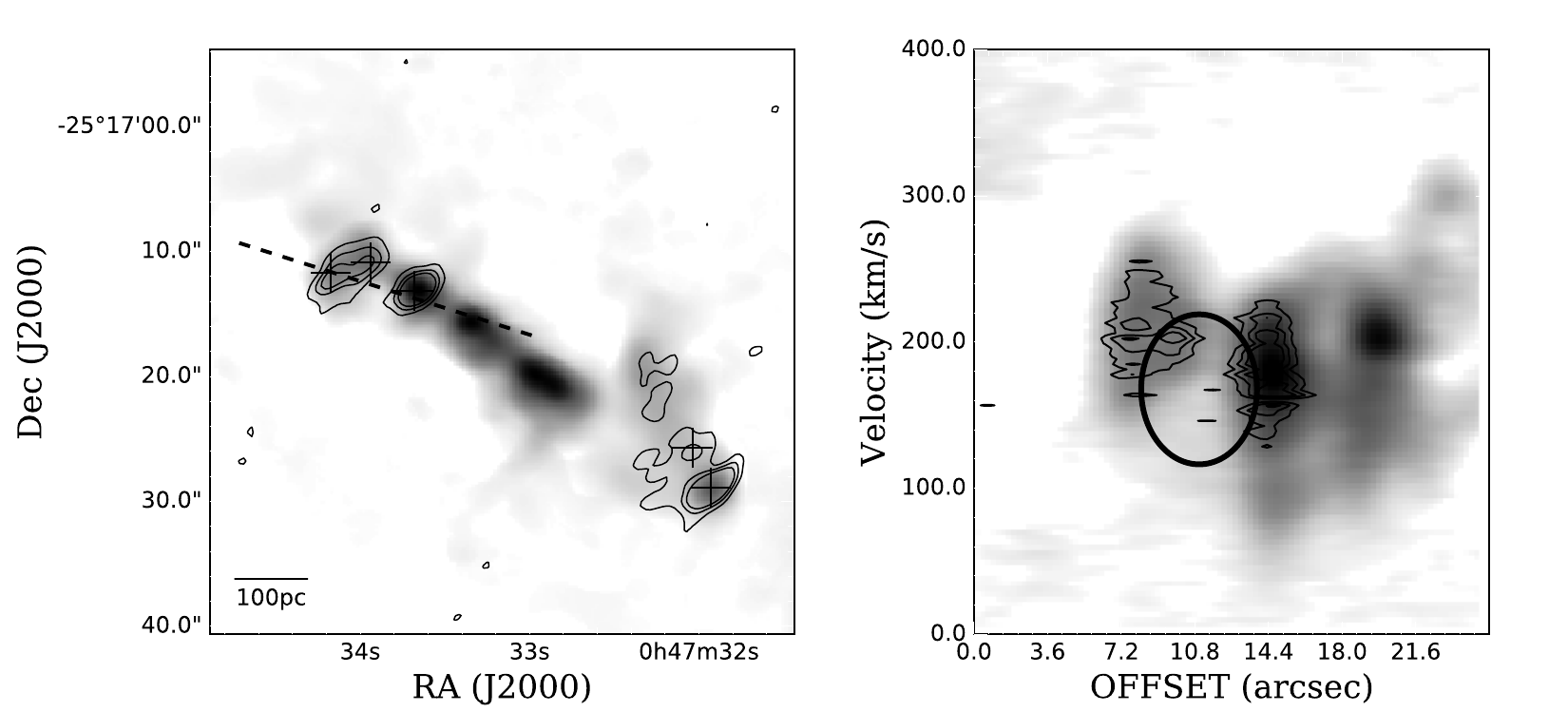}
\includegraphics[width=0.95\textwidth]{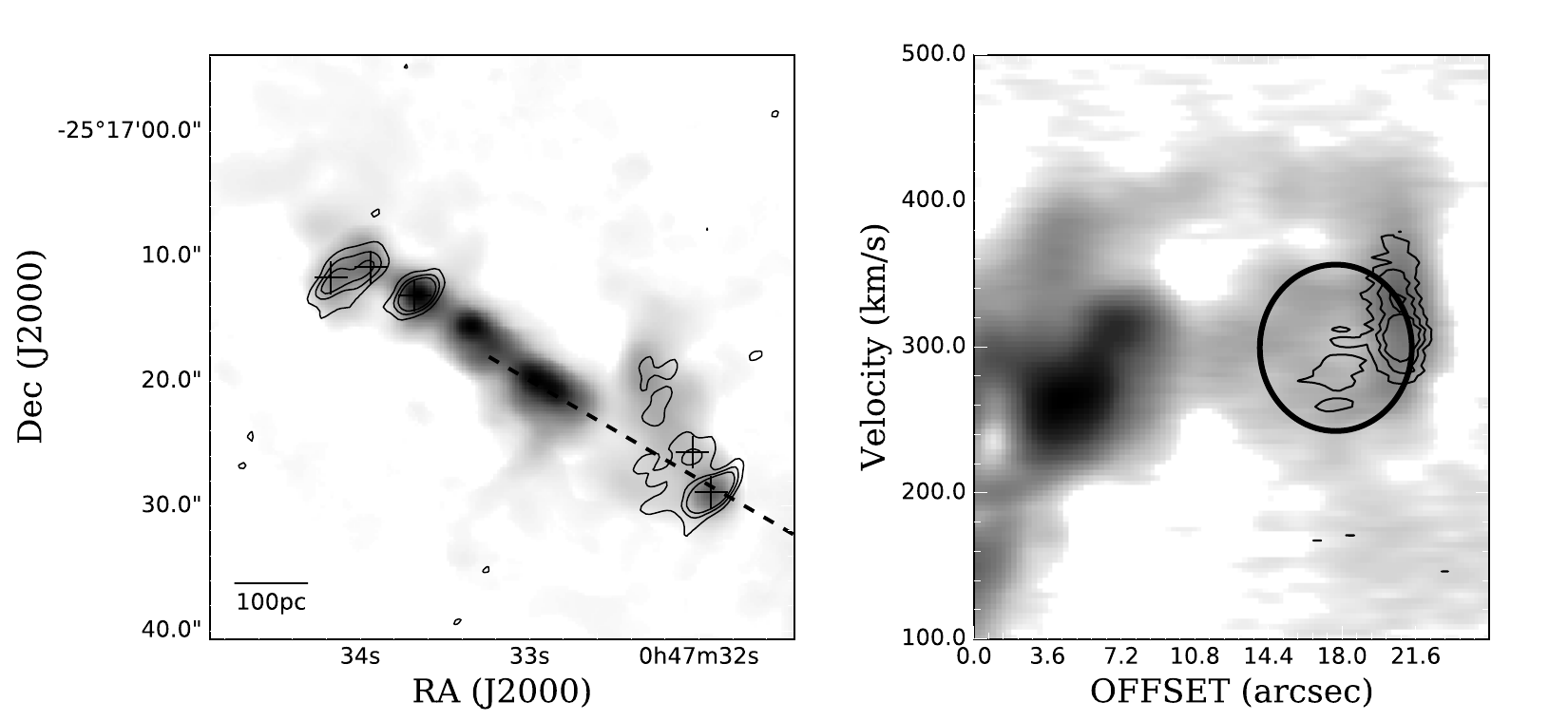}
\caption{Position-Velocity (PV) cuts through the HCN  (grayscale) and  36 GHz methanol maser (3, 6, and 9$\sigma$ contours shown in black) data cubes. The PV cuts show a few shell-like structures marked with black circles. The methanol masers appear to be concentrated around the edges of the shells. }
\label{fig:pvcuts}
\end{figure*}

At the same time, we measure a relative decrement of HNCO in the center compared to the locations with 36 GHz \methanol\ masers and the expanding \citet{Saka2011} superbubbles.
 In the case of NGC 253 \citet{Meier2015} argue that the relative decrement of HNCO is a consequence of dissociation in the PDR dominated center. The lack of methanol masers on the eastern side of the west superbubble is consistent with dissociation closer to the center.  It is interesting that the non-thermal 36 GHz \methanol\ correlates with the HNCO. A similar tight correlation has been observed between HNCO and thermal \methanol\ in nuclei(e.g. \citealt{MT2005}, \citealt{Meier2012}). 
This is further evidence that  that the 36 GHz \methanol\ is tracing similar shocked regions. SiO is also enhanced near the superbubbles. However SiO is more uniformly distributed in the center 300 pc of NGC 253 where we did not observe any 36 GHz \methanol\ masers suggesting that the 36 GHz \methanol\ maser is more closely related to weak shocks than strong shocks.  Position velocity (PV) cuts were made through the eastern and western groups of masers. Figure \ref{fig:pvcuts} shows the PV cuts through the HCN cube with the naturally weighted 36 GHz methanol contours in black. The HCN cube reveals a few shell-like structures. The methanol masers are well correlated with the edges of the shells strengthening the connection between the shocks and the dense molecular gas. 

\section{Summary}
We have provided analysis of VLA observations of the dense gas and masers associated with the nuclear starburst in NGC 253. We conclude the following:
\begin{enumerate}

\item We have detected \amm(1,1), (2,2), (3,3), (4,4), and (5,5) in the central kpc of NGC 253. \amm(9,9) is detected in only one location. No \amm\ is observed in the molecular outflow.  

\item We find the molecular gas in NGC 253 to be best described by a spatially uniform two kinetic temperature model with a warm 130 K component and a cool 57 K component. The continuum peak (C1) may indeed be hotter as it is closer to the starburst, but absorption of the metastable para-\amm\ lines make temperature measurements less accurate. Direct LVG analysis corroborates a two temperature model of NGC 253, though with a cool component of 74 K and a warm component of 145K. The LVG analysis does hint at a hotter component greater than 300 K as traced by the \amm(4,4) and \amm(5,5) ratio. The temperature distribution is not well correlated with PDRs, weak shocks, or strong shocks, and is constant over the central kpc. 

\item We confirm the result from \citet{Ott2005} suggesting that there exist \amm(3,3) masers in the nucleus of NGC 253. There is currently only one other known source of extragalactic \amm(3,3) masers,  NGC 3079 \citep{Miyamoto2015}. Both galaxies host outflows, but in NGC 253 the outflow is driven by a starburst whereas in NGC 3079 it is likely driven by an AGN.

\item Expanding superbubbles do not appear to heat the gas in NGC 253. A comparison with the results from \citet{leroy2015} shows that the linewidths are dominated by turbulent motions within a GMC. The large \amm\ linewidths are likely due to the bulk motion of the GMCs. 

\item Three \water\ maser features have been observed. The \water\ maser W1 is coincident with the continuum peak. 
We show indications emission is extended along the minor axis of the galaxy, the spectrum suggests multiple components and the velocities are more similar to the outflow than the disk. It is likely that this maser is related to the bipolar outflow of ionized gas, though other explanations are possible.  
High resolution followup observations are needed to confirm a relationship with the bipolar outflow.

W2 is located to the southwest of W1 and shows multiple spectrally unresolved masers with velocities centered about systemic, suggesting these masers exist in a circumnuclear torus.  It is also possible that  W2  may in part trace the receding side of the outflow.  W3 is unresolved and located to the southwest of W1 and W2. Its progenitor is likely a massive star.  

\item We detect five regions with 36 GHz \methanol\ masers at the outside edge of the central kpc as traced by \amm. The spatial proximity and velocities suggest a relationship with known superbubbles. Position velocity cuts  show that all the 36 GHz methanol masers may be related to shell-like features.   The 36 GHz \methanol\ morphology and HNCO are similar. This similarity suggests that both HNCO and 36 GHz methanol masers are tracing similar conditions. 

\end{enumerate}

Our analysis reveals the properties of the dense molecular ISM in along the central kpc of NGC 253 using centimeter and millimeter emission and absorption lines. We have uncovered relationships with the bipolar outflow, expanding molecular bubbles, and the nuclear starburst. 

\acknowledgments 
Mark Gorski acknowledges support from the National Radio Astronomy Observatory in the form of a graduate student internship, and a Reber Fellowship. We would like to thank Alberto Bolatto and Adam Leroy for sharing the ALMA image cubes. This research has made use of the NASA/IPAC Extragalactic Database (NED), which is maintained by the Jet Propulsion Laboratory, Caltech, under contract with the National Aeronautics and Space Administration (NASA) and NASA's Astrophysical Data System Abstract Service (ADS).

\clearpage


\begin{deluxetable*}{llllllllll}
\tabletypesize{\footnotesize}
\tablewidth{0pt}
\tablecolumns{8}
\tablecaption{\amm\ Line Parameters From A1-A7\label{tab:ammprop}}
\tablehead{ 
\colhead{Location} & \colhead{A1} & \colhead{A2 }& \colhead{A3} & \colhead{A4} & \colhead{A5} & \colhead{A6} & \colhead{A7} \\
RA (J2000) {hh}:{mm}:{ss} &\phn00:47:34.0	&\phn00:47:33.6	&\phn00:47:33.3	&\phn00:47:32.8	&\phn00:47:32.3	&\phn00:47:32.2	&\phn00:47:31.9\\
DEC (J2000) \phn{\arcdeg}~\phn{\arcmin}~\phn{\arcsec} &-25 17 11.2	&-25 17 12.8	&-25 17 15.3	&-25 17 21.1	&-25 17 19.9	&-25 17 25.2	&-25 17 28.7\\
Distance from C1* (pc) & \colhead{202} & \colhead{131} & \colhead{53} & \colhead{107} & \colhead{208} & \colhead{253} & \colhead{342}
}
\startdata
\cutinhead{ \amm (1,1) }
$\int T_{mb} d\nu$ (K \kms)	& \phn48.8$\pm$2.1		&\phn66.8$\pm$2.4		& \phn24.6$\pm$2.0		& \phn57.2$\pm$2.2 		& \phn61.0$\pm$3.0		& \phn62.4$\pm$2.6		& \phn59.3$\pm$2.3	\\
$V_{LSRK}$ (\kms) 			& 200.1$\pm$1.8		& 172.8$\pm$1.4		& \phn62.8$\pm$2.1		& 278.2$\pm$0.1		& 313.2$\pm$2.3		& 294.5$\pm$1.9		& 299.6$\pm$1.4	\\
$V_{FWHM}$ (\kms) 		& \phn83.6$\pm$4.2 		& \phn80.8$\pm$3.5		& \phn54.2$\pm$5.0		& \phn65.3$\pm$3.2		& \phn95.2$\pm$6.0		& \phn90.0$\pm$4.3		& \phn74.9$\pm$3.5	\\
$T_{mb}$ (K) 				& \phn0.55$\pm$0.08	& \phn0.78$\pm$0.09	& \phn0.43$\pm$0.09	& \phn0.81$\pm$0.09	& \phn0.60$\pm$0.10	& \phn0.65$\pm$0.09	& \phn0.74$\pm$0.09	\\\cutinhead{ \amm (2,2) }
$\int T_{mb} d\nu$ (K \kms)	& \phn44.4$\pm$2.1 		& \phn49.8$\pm$1.9		& \phn16.7$\pm$1.7		& \phn42.4$\pm$1.7		& \phn48.3$\pm$2.1		& \phn47.0$\pm$2.3		& \phn39.3$\pm$2.0		\\
$V_{LSRK}$ (\kms) 			& 225.4$\pm$2.0		& 198.6$\pm$1.4		& 192.2$\pm$2.6		& 309.5$\pm$1.1		& 336.4$\pm$2.0		& 323.2$\pm$2.3		& 328.5$\pm$1.8	 \\
$V_{FWHM}$ (\kms) 		& \phn87.1$\pm$5.3		& \phn73.6$\pm$3.3		& \phn49.7$\pm$5.6		& \phn55.1$\pm$2.6		& \phn89.1$\pm$4.6		& \phn91.4$\pm$5.1 		& \phn74.3$\pm$4.4	\\
$T_{mb}$ (K) 				& \phn0.45$\pm$0.08	& \phn0.64$\pm$0.08	& \phn0.32$\pm$0.09	& \phn0.72$\pm$0.08	& \phn0.51$\pm$0.08	& \phn0.48$\pm$0.08	& \phn0.50$\pm$0.08	\\
\cutinhead{ \amm (3,3) }
$\int T_{mb} d\nu$ (K \kms)	& \phn74.0$\pm$12.1	& \phn132.0$\pm$2.3	& \phn125.5$\pm$3.2	& \phn126.9$\pm$2.3	& \phn83.1$\pm$2.4		& \phn89.9$\pm$2.5		& \phn80.9$\pm$1.6 \\
$V_{LSRK}$ (\kms) 			& 204.9$\pm$0.8		& 176.6$\pm$0.6		& 176.2$\pm$0.8		& 283.9$\pm$0.5		& 316.6$\pm$1.4		& 287.9$\pm$1.3		& 308.5$\pm$0.6 \\
$V_{FWHM}$ (\kms) 		& \phn68.8$\pm$2.0		& \phn74.4$\pm$1.4		& \phn63.7$\pm$2.0		& \phn61.5$\pm$1.4		& \phn103.5$\pm$3.7	& \phn93.7$\pm$3.1		& \phn66.1$\pm$1.6 \\
$T_{mb}$ (K) 				& \phn1.01$\pm$0.07	& \phn1.39$\pm$0.08	& \phn1.85$\pm$0.14	& \phn1.94$\pm$0.10	& \phn0.78$\pm$0.08	& \phn0.90$\pm$0.09	& \phn1.15$\pm$0.07 \\
\cutinhead{ \amm (4,4) }
$\int T_{mb} d\nu$ (K \kms)	& \phn15.0$\pm$2.1		& \phn23.9$\pm$1.9		& \phn\phn8.0$\pm$1.4	& \phn19.8$\pm$1.3		& \phn24.1$\pm$2.3		& \phn20.0$\pm$2.2		& \phn20.1$\pm$1.7	\\
$V_{LSRK}$ (\kms) 			& 199.0$\pm$3.7		& 176.9$\pm$2.8		& 166.5$\pm$3.6		& 286.7$\pm$1.6		& 301.9$\pm$5.1		& 300.3$\pm$4.5		& 310.0$\pm$2.8	\\
$V_{FWHM}$ (\kms) 		& \phn59.5$\pm$8.6		& \phn70.5$\pm$6.4		& \phn36.5$\pm$6.6		& \phn44.5$\pm$3.3		& \phn125.4$\pm$13.7	&\phn82.4$\pm$10.4		& \phn64.8$\pm$6.2	 \\
$T_{mb}$ (K) 				& \phn0.24$\pm$0.09	& \phn0.32$\pm$0.08	& \phn0.21$\pm$0.08	& \phn0.41$\pm$0.07	& \phn0.17$\pm$0.07	& \phn0.23$\pm$0.08	& \phn0.29$\pm$0.09	\\
\cutinhead{ \amm (5,5) }
$\int T_{mb} d\nu$ (K \kms)	& \phn12.1$\pm$2.4		& \phn11.0$\pm$1.3		& \phn\phn5.2$\pm$1.3	& \phn16.9$\pm$1.8		& \phn15.0$\pm$2.5		& \phn13.2$\pm$2.2		& \phn11.2$\pm$1.5		 \\
$V_{LSRK}$ (\kms) 			& 195.9$\pm$6.1		& 164.5$\pm$2.5		& 168.3$\pm$3.7		& 281.4$\pm$2.7		& 316.1$\pm$7.3		& 285.6$\pm$8.3		& 294.2$\pm$2.4	 \\
$V_{FWHM}$ (\kms) 		& \phn71.5$\pm$19.2	& \phn40.9$\pm$5.2		& \phn29.0$\pm$7.7		& \phn53.9$\pm$7.6		& \phn91.2$\pm$19.2	& \phn103.5$\pm$19.6	& \phn36.5$\pm$5.1	 \\
$T_{mb}$ (K) 				& \phn0.15$\pm$0.08	& \phn0.25$\pm$0.08	& \phn0.17$\pm$0.09	& \phn0.29$\pm$0.08	& \phn0.15$\pm$0.09	& \phn0.12$\pm$0.07	& \phn0.28$\pm$0.09	\\
\cutinhead{ Rotation Temperatures  $T_{J J\prime}$}
$T_{12}$	(K)				& $ 46^{+5}_{-4} $		& $37^{+2}_{-2}$		& $34^{+6}_{-4}	$		& $37^{+3}_{-3}$		& $39^{+4}_{-3}$		& $38^{+3}_{-3}$		& $34^{+2}_{-2}$\\
$T_{24}$	(K)				& $72^{+7}_{-6} $		& $89^{+7}_{-6}$		& $89^{+18}_{-13}	$	& $87^{+6}_{-6}$		& $91^{+9}_{-8}$		& $82^{+8}_{-8}$		& $93^{+9}_{-8}$\\
$T_{45}$	(K)				& $202^{+403}_{-84}$ 	& $91^{+21}_{-15}$		& $143^{+213}_{-55}$	& $226^{+151}_{-67}$	& $198^{+69}_{-35}$	& $140^{+95}_{-42}$	& $112^{+37}_{-23}$\\
\cutinhead{ Kinetic Temperatures  $T_{KinJ J\prime}$}
$T_{Kin12}$	(K)			& $ 99^{+38}_{-23} $		& $60^{+11}_{-8}$		& $50^{+23}_{-12}	$	& $59^{+11}_{-8}$		& $68^{+19}_{-12}$		& $62^{+15}_{-10}$		& $48^{+9}_{-6}$\\
$T_{Kin24}$	(K)			& $107^{+15}_{-12} $	& $148^{+22}_{-11}$		& $147^{+62}_{-34}	$	& $143^{+18}_{-15}$		& $154^{+30}_{-22}$		& $130^{+23}_{-17}$		& $158^{+31}_{-23}$\\
$T_{Kin45}$	(K)			& $229^{+460}_{-96}$ 	& $103^{+24}_{-17}$		& $162^{+243}_{-64}$	& $257^{+173}_{-76}$	& $145^{+79}_{-40}$		& $159^{+108}_{-48}$	& $126^{+43}_{-27}$\\
\enddata
\tablenotetext{*}{Right Ascension 00h 47m 33.160s Declination -25$^\circ$ 17' 17.118\arcsec  }
\end{deluxetable*}


\begin{deluxetable*}{lllll}
\tabletypesize{\footnotesize}
\tablewidth{0pt}
\tablecolumns{5}
\tablecaption {Line Parameters Towards C1\label{tab:C1prop}}
\tablehead{
\amm & $\int T_{mb} d\nu$ 	& $V_{LSRK}$ 	& $V_{FWHM}$ 	& $T_{mb}$    \\
(J,K)& (K \kms)			& (\kms)		& (\kms)			&(K)
}
\startdata
\cutinhead{Absorption}
\amm(1,1)	&-8.1$\pm$1.5		&	223.3$\pm$3.4 &	32.3$\pm$6.9 	& -0.23$\pm$0.10 \\
\amm(2,2)	&-17.5$\pm$1.7 	&	247.7$\pm$2.6 &	49.7$\pm$6.6	& -0.33$\pm$0.08 \\
\amm(4,4)	&-10.7$\pm$2.4 	&	223.8$\pm$6.9 &	60.0$\pm$16.3 	& -0.17$\pm$0.11 \\
\amm(5,5)	&-6.8$\pm$1.3 		&	218.7$\pm$2.9 &	29.4$\pm$6.1 	& -0.22$\pm$0.09 \\
\cutinhead{Emission}
\amm(3,3)a	&48.6$\pm$5.1 &	171.1$\pm$1.1 &	55.1$\pm$3.2 &	0.82$\pm$0.09 \\
\amm(3,3)b	&86.7$\pm$6.2 &	256.6$\pm$4.5 &	129.6$\pm$9.5 &	0.63$\pm$0.09 \\
\enddata
\end{deluxetable*}


\begin{deluxetable*}{lll}
\tabletypesize{\footnotesize}
\tablewidth{0pt}
\tablecolumns{3}
\tablecaption{\amm\ (9,9) Line Parameters at A3 \label{99prop}}
\tablehead{ &
} 
\startdata
$\int T_{mb} d\nu$ (K \kms)	& \phn\phn6.9$\pm$1.2	 		\\
$V_{LSRK}$ (\kms)			& 164.7$\pm$3.5		\\
$V_{FWHM}$ (\kms)			&\phn38.1$\pm$6.7		\\
$T_{mb}$ (K) 				&\phn0.17$\pm$0.07	
\enddata
\end{deluxetable*}


\begin{deluxetable*}{llclllll}
\tabletypesize{\footnotesize}
\tablewidth{0pt}
\tablecolumns{8}
\tablecaption{\water\ Maser Candidates\label{tab:waterprop}}
\tablehead{
RA (J2000)& Dec(J2000) &
Velocity Component & $\int T_{mb} d\nu$ 	& $V_{LSRK}$ 	& $V_{FWHM}$ 	& $T_{mb}$  & Luminosity  \\
{hh}:{mm}:{ss}&\phn{\arcdeg}~\phn{\arcmin}~\phn{\arcsec} & & (K \kms)			& (\kms)		& (\kms)			&(K) & $L_\odot$ 
}
\startdata
\cutinhead{W1 }
00:47:33.1	&-25 17 16.9	&a		&214.9$\pm$3.5			&	109.1$\pm$0.3 		&	42.0$\pm$0.9 			& 4.7$\pm$0.2		&0.667 	\\
			&			&b		&\phn\phn6.2$\pm$1.1 		&	\phn27.0$\pm$0.4 	&	\phn3.5$\pm$0.5		& 1.7$\pm$0.2		&0.019	\\
			&			&c		&\phn\phn5.5$\pm$0.1		&	\phn21.0			&	\phn3.5$^*$ 				& 1.6$\pm$0.2 		&0.017	\\
\cutinhead{W2}
00:47:33.0	&-25 17 19.0	&		&25.7$\pm$2.7		&	233.6			&	107.2$\pm$14.0				&0.2$\pm$0.09		&0.080	\\
\cutinhead{W3}
00:47:32.8	&-25 17 20.7	&		&7.4$\pm$0.5		&	303.6$\pm$1.4 		&	3.5$\pm$1.1 		& 2.01$\pm$0.10		&0.023
\enddata
\tablenotetext{*}{unresolved, V$_{FWHM}$ is an upper limit }
\end{deluxetable*}


\begin{deluxetable*}{llclllll}
\tabletypesize{\footnotesize}
\tablewidth{0pt}
\tablecolumns{8}
\tablecaption{\methanol\ Maser Candidates\label{tab:methanolprop}}
\tablehead{
RA(J2000)& Dec(J2000) &  Velocity Component & $\int T_{mb} d\nu$ 	& $V_{LSRK}$ 	& $V_{FWHM}$ 	& $T_{mb}$  & Luminosity  \\
{hh}:{mm}:{ss}&\phn{\arcdeg}~\phn{\arcmin}~\phn{\arcsec} & & (K \kms)			& (\kms)		& (\kms)			&(K) & $L_\odot$ 
}
\startdata
\cutinhead{M1}
00:47:34.1	& -25 17 11.7	&	&1.38$\pm$0.07	&	202.0$\pm$0.3	&	49.2$\pm$3.1	& 0.026$\pm$0.005	&0.63 \\
\cutinhead{M2}
00:47:33.9	& -25 17 10.8 	&	&1.41$\pm$0.07	&	195.4$\pm$0.8	&	32.2$\pm$2.3	&0.040$\pm$0.006	&0.65 \\
\cutinhead{M3}
00:47:33.7	& -25 17 13.1	&	&3.59$\pm$0.14	&	165.9$\pm$1.7	&	85.8$\pm$4.4	&0.039$\pm$0.008	&1.63 \\
\cutinhead{M4}
00:47:31.9	&-25 17 28.9	&	&3.09$\pm$0.18	&	294.1$\pm$2.6	&	84.3$\pm$5.6	&0.035$\pm$0.005	&1.42 \\
\cutinhead{M5}
00:47:32.0	&-25 17 25.7	&a	&2.19$\pm$0.89	&	294.3$\pm$3.0	&	35.7$\pm$4.8	&0.056$\pm$0.005 	&1.01\\
			&			&b	&2.33$\pm$0.96	&	331.6$\pm$10.2&	54.5$\pm$16.9	&0.041$\pm$0.005 	&1.06\\
\enddata
\end{deluxetable*}


\begin{deluxetable*}{lllllllll}
\tabletypesize{\footnotesize}
\centering
\tablewidth{0pt}
\tablecolumns{8}
\tablecaption{\amm\ LVG Best Fit Values \label{tab:LVG}}
\tablehead{ 
\colhead{Location} & \colhead{A1} & \colhead{A2 }& \colhead{A3} & \colhead{A4} & \colhead{A5} & \colhead{A6} & \colhead{A7} }
\startdata
\cutinhead{\amm(1,1)/\amm(2,2)}
$T_{Kin}$ (K) 				&117	$\pm$135	&69$\pm$24	 &54$\pm$27	&69$\pm$24	 &81$\pm$41 		&72$\pm$	27	&57$\pm$	18\\
\cutinhead{\amm(2,2)/\amm(4,4)}
$T_{Kin}$ (K) 				&111$\pm$24	&156$\pm$39 	&165$\pm$75	&150$\pm$33	&165$\pm$48 	&108$\pm$24 	&162$\pm$42	\\
\enddata
\tablecomments{This table reports the median temperatures of the best fit LVG model to the data.  The uncertainty on the kinetic temperature fit are asymmetric, therefore uncertainty reported here is the greater $1\sigma$ deviation from the best fit.}
\end{deluxetable*}


\end{document}